\documentclass[graybox,natbib,nosecnum]{svmult}
\bibpunct{(}{)}{;}{a}{}{,} 

\pdfoutput=1   

\usepackage{mathptmx}       
\usepackage{helvet}         
\usepackage{courier}        
\usepackage{type1cm}        

\usepackage{makeidx}         
\usepackage{graphicx}        
\usepackage{multicol}        
\usepackage[bottom]{footmisc}
\usepackage[normalem]{ulem}	
\usepackage{hyperref}  

\usepackage{soul}   

\newcommand*\aap{A\&A}

\newcommand*\aj{AJ}

\newcommand*\apj{ApJ}
\newcommand*\apjl{ApJ}

\newcommand*\araa{ARA\&A}

\newcommand*\icarus{Icarus}

\newcommand*\mnras{MNRAS}

\newcommand*\nat{Nature}

\newcommand*\pasp{PASP}

\newcommand{\MSun}{ M_\odot}

\newcommand{\Mjup}{ M_{\rm J}}

\newcommand{\Mearth}{ M_\oplus}

\newcommand{\Op}{\Omega_{\rm p}}
\newcommand{\Os}{\Omega_{\ast}}

\usepackage[normalem]{ulem}
\usepackage{color}
\definecolor{blue}{RGB}{0,0,255}
\definecolor{red}{RGB}{255,0,0}
\definecolor{green}{RGB}{0,200,0}
\definecolor{black}{RGB}{0,0,0}

\makeindex             

\begin{document}

\title*{Habitability in brown dwarf systems}
\author{Emeline Bolmont}
\institute{Laboratoire AIM Paris-Saclay, CEA/Irfu Universit\'e Paris-Diderot CNRS/INSU, 91191 Gif-sur-Yvette, France, \email{emeline.bolmont@cea.fr}}
%
%
\maketitle

\abstract{The very recent discovery of planets orbiting very low mass stars sheds light on these exotic objects. 
Planetary systems around low-mass stars and brown dwarfs are very different from our solar system: the planets are expected to be much closer than Mercury, in a layout that could resemble the system of Jupiter and its moons. 
The recent discoveries point in that direction with, for example, the system of Kepler-42 and especially the system of TRAPPIST-1 which has seven planets in a configuration very close to the moons of Jupiter.
Low-mass stars and brown dwarfs are thought to be very common in our neighborhood and are thought to host many planetary systems.
The planets orbiting in the habitable zone of brown dwarfs (and very low-mass stars) represent one of the next challenges of the following decades: they are the only planets of the habitable zone whose atmosphere we will be able to probe (e.g. with the JWST).}

\section{Introduction }

In the 10 pc around us, 75\% of objects are low mass stars and brown dwarfs (RECONS project \url{http://www.recons.org/ }, e.g. \citealt{Henry2016}). 
It is thought that a majority of those low-mass stars and brown dwarfs (BDs) host planetary systems \citep[e.g.][]{DressingCharbonneau2015}. 
For instance, the population of Earth-size planets in the habitable zone (HZ) of low-mass stars has been estimated to be between $\sim20$\% \citep{DressingCharbonneau2015} and $\sim40-50$\% \citep{Bonfils2013, DressingCharbonneau2013, Kopparapu2013}. 
The HZ is here defined as the region around a star in which a planet with a sufficiently dense atmosphere can host surface liquid water \citep[e.g.][]{Kasting1993,Selsis2007}. 

Because of the low luminosity of these objects, planets inside the HZ are sufficiently close-in to be influenced by the tidal interactions between the dwarf and the planets \citep{Barnes2008, Barnes2009, 	Barnes2010ASPC, Barnes2011ASPC}.
Moreover, the spectral distribution of these dwarfs and the proximity of the HZ would likely cause the climate to be very different from that of the Earth \citep[e.g.,][]{Segura2005, Rauer2011}.

This situation is even more extreme for BDs. 
They are not massive enough to start the hydrogen fusion reaction \citep{ChabrierBaraffe1997, ChabrierBaraffe2000} so their temperature is even cooler than for M-dwarfs and they also cool down with time. 
Their HZ therefore moves inward and can even be within the Roche limit at late ages. 
Planets in the HZ of brown dwarfs should thus be submitted to strong tides \citep{Bolmont2011} which influence their orbit and rotation.

The existence and fate of planets around BDs has been considered punctually over the years \citep[e.g., the works of][]{Desidera1999, Andreeshchev2004, Bolmont2011, BarnesHeller2013}, but the recent discoveries of planetary systems around very low-mass stars (Kepler-42, a $0.13~\MSun$ dwarf with at least three small planets: \citealt{Muirhead2012}; Proxima, a $0.123~\MSun$ dwarf hosting at least a planet: \citealt{Anglada-Escude2016}; and TRAPPIST-1, a $\sim0.08~\MSun$ dwarf with a multiple planet system: \citealt{Gillon2016, Gillon2017}) have contributed to renew the interest on these objects.

The discovery of the TRAPPIST-1 planets also illustrates the importance to study those objects: indeed those planets are the only known planets of the HZ for which we will be able to probe the atmosphere with future instruments such as the JWST \citep[e.g.][]{Belu2013} or the E-ELT \citep{Rodler2014}.
Note that other planets around low-mass stars could be targets for the JWST, like LHS~1140b \citep{Dittmann2017}. 
However due to its high surface gravity and the scarcity of its transits, it would be a much more technically challenging observation than that of the TRAPPIST-1 planets.
Note that TRAPPIST-1 is particularly interesting in the framework of this chapter, because its estimated mass is just above the theoretical limit between BDs and low-mass stars.

We first introduce brown dwarfs and discuss why they are so special in terms of hosting environments for planets and then explore different aspects of the habitability of planets orbiting brown dwarfs.

\section{Brown dwarfs and their evolution}

Before ever actually observing BDs, the existence of these objects has been conceptualized by \citet{Kumar1963} and \citet{Hayashi1963} as being failed stars: not massive enough for the core temperature to reach values allowing the fusion of hydrogen.
We first give the definition of a BD and then discuss what makes them so special when considering the habitability of potential planets.

\subsection{What are brown dwarfs?}

BDs are objects which are thought to form like stars \citep{Luhman2012}, by the gravitational collapse of a molecular cloud (thermal radio jets, typical of young stellar systems, have been detected in young BDs systems. This shows the continuity of the formation processes between low mass stars and BDs; e.g. \citealt{Morata2015}). 
However, contrary to stars, these objects are not massive enough for their core temperature to reach the level needed to initiate the hydrogen fusion nuclear reaction (see \citealt{Luhman2012} for a review on BDs).
They are therefore very faint and were first observed quite recently by \citet{Nakajima1995} and \citet{Rebolo1995}.

As the fusion temperature of hydrogen is of about $3\times10^6$~K, only objects of more than $M_\star\approx 75~\Mjup$ can initiate the PPI fusion reaction chain.
This gives an upper mass for BDs \citep{ChabrierBaraffe1997, ChabrierBaraffe2000}.
The lower limit is however much less well defined. 
There are some arguments both observational \citep{Caballero2007} and analytical \citep{PadoanNordlund2004, HennebelleChabrier2008} which suggest that the same star formation process can produce objects down to a few mass of Jupiter.
One possible lower limit definition would be the deuterium fusion limit: studies show that objects of mass higher than $\approx 13~\Mjup$ can still initiate the deuterium fusion reaction while objects less massive cannot.
BDs would therefore be objects in the mass range of $13$ to $75~\Mjup$ and all objects with a mass lower than $13~\Mjup$ would be planets \citep[IAU definition, see][]{Boss2007}. 
However this definition is more of an indication rather than a strong astrophysical limit. 
Indeed, \citet{Spiegel2011} showed that the limit of $13~\Mjup$ can change between $11$ and $16~\Mjup$ when considering different metallicities for example.

The population of \textit{mini brown dwarfs} and \textit{giant planets} (formed in a protoplanetary disk) can have a common mass range.
It is therefore fundamental to try to differentiate those two types of astrophysical objects  \citep{Leconte2009, Spiegel2011, Leconte2011a}.

\bigskip

Since the discovery of the first BDs in 1995, many more have been detected in star forming regions (in the Chamaeleon I cloud: \citet{Comeron2000} and \citet{LopezMarti2004}), in open clusters (e.g., in the Pleiades:  \citet{Zapatero1997}, and in the young open cluster IC 2391: \citet{Barrado2001}) and also among field objects \citep{Kirkpatrick1999,Phan-Bao2001}.

The number of detected BDs have thus risen thanks to observation missions such as 2MASS (\textit{Two Micron All Sky Survey}, \citealt{Skrutskie2006}) or WISE (\textit{Wide-field Infrared Survey Explorer}, \citealt{Cushing2011}).
Due to their low temperature, BDs emit principally in the infrared and are therefore detected by instruments probing these wavelengths.
As of 2016, the number of BDs (objets of spectral type L, T and Y) is of about 2800 (\url{http://www.johnstonsarchive.net/astro/browndwarflist.html}).

Studies based on the Initial Mass Function \citep[e.g.][]{Salpeter1955} tend to show that the number of low mass objects like BDs should be much higher than more massive stars.
\citet{Chabrier2002} revisited these studies focusing on BDs and showed that there should be as many BDs as there are stars.
However, thanks to the intensive observational efforts, it has been shown that brown dwarfs are more scarce than previously thought.
For example, the WISE survey \citep{Kirkpatrick2013} showed that within 8~parsec, there are 33 BDs and 211 stars (white dwarfs, O, B, A, F, G, K, M stars), which yield that there is 1 BD for every 6.4 stars (interestingly, there are 4.1 BDs for every G-star).
More recently, the RECONS team \citep{Henry2016} confirmed this tendency showing that there is one BD for every $\sim10$~stars within 10~parsec.

\subsection{What is special about them?}

\begin{figure}
	\begin{center}
	\includegraphics[width=9cm]{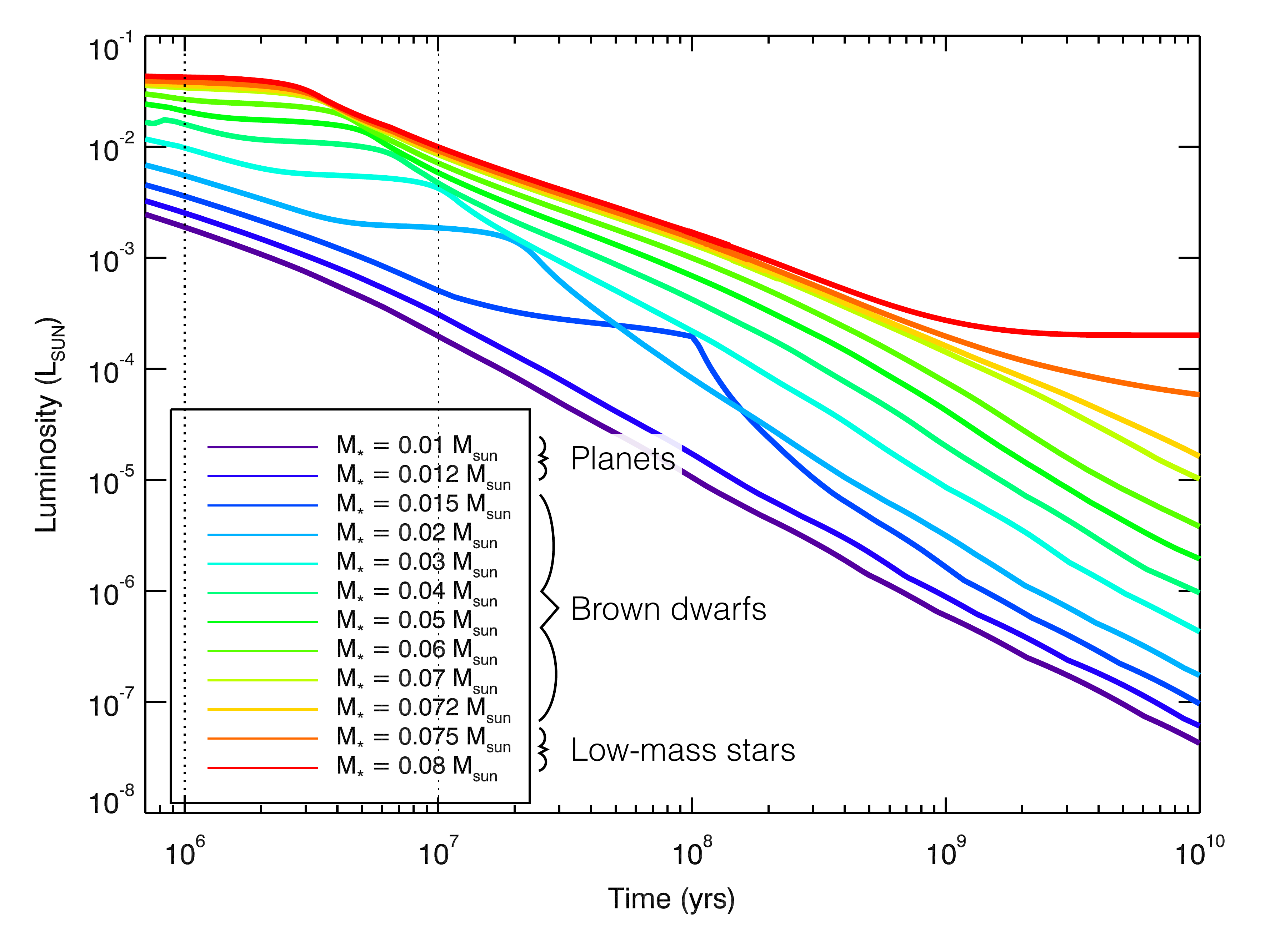}
	\end{center}
	\caption{Evolution of the luminosity of low-mass objects of different masses \citep[grids coming from][]{Leconte2011}. The two lower mass objects are planets in this model (the limit of deuterium fusion is never reached). The two upper mass objects are low-mass stars (the deuterium is quickly burnt in a few Myr, and the hydrogen fusion begins after a few Gyr). Figure adapted from \citet{phd_Bolmont2013}.}
	\label{bolmont_fig1}
	\end{figure}

Because BDs cannot initiate the hydrogen fusion reaction in their core, the energy due to this reaction is here missing to prevent the contraction and the cooling down of these objets.
However, by definition, BD are massive enough to initiate the deuterium fusion. 
As this additional source of energy is able to compensate the radiative losses for a while, the contraction is slowed down and radius and luminosity reach a plateau.
As the primordial abundance of deuterium is small and the reaction constants are big, this phase lasts only a few million years for massive BDs and about $100$~million years when they are close to the deuterium fusion limit.
Fig.~\ref{bolmont_fig1} shows the evolution of the luminosity of several low-mass objects: from planets of $0.01~\MSun$ ($\sim 10~\Mjup$) to low-mass stars of $0.08~\MSun$ passing by BDs. 

	\begin{figure}
	\begin{center}
	\includegraphics[width=9cm]{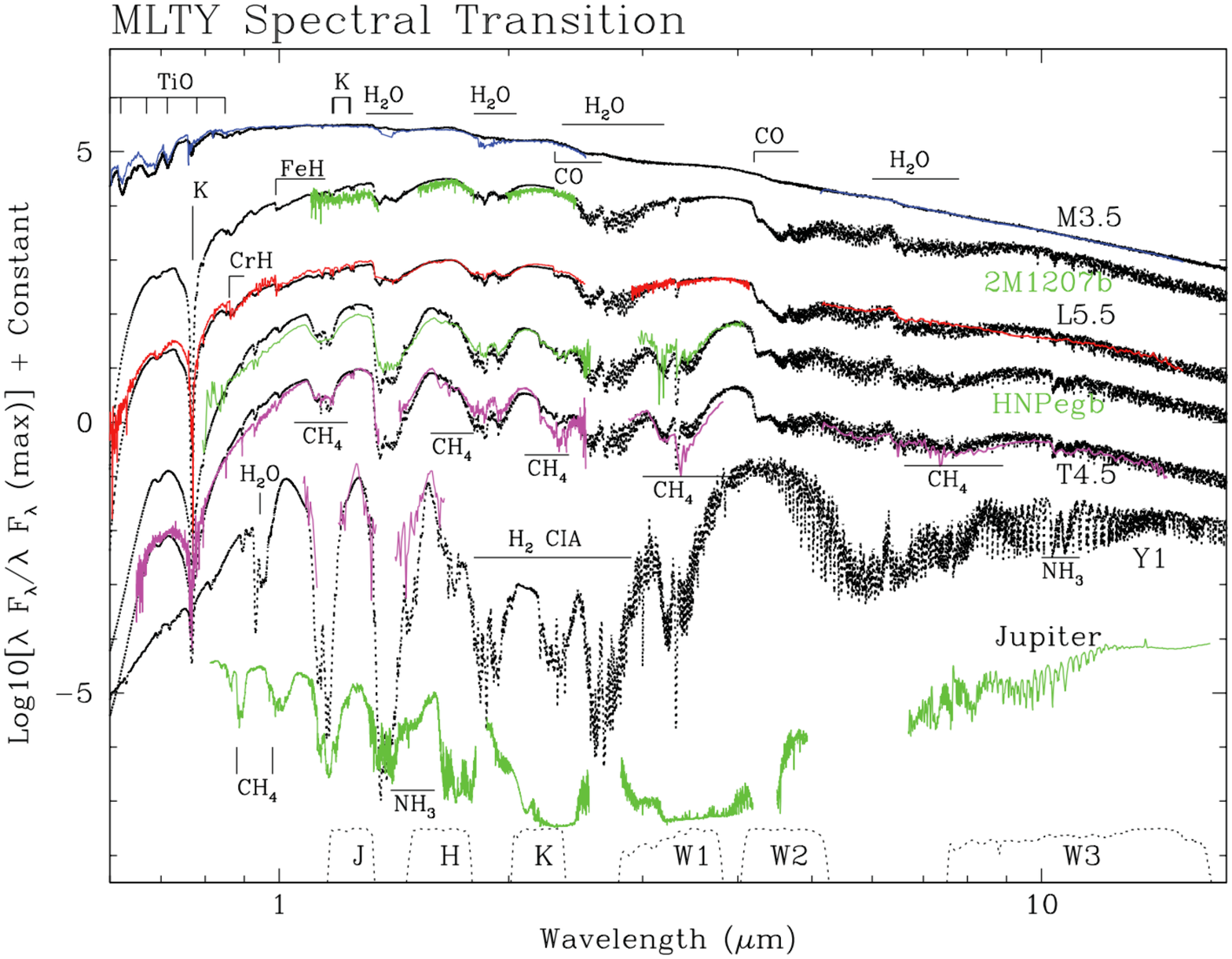}
	\end{center}
	\caption{Spectral Energy Distribution of dwarfs of different spectral types. The full lines represent measurements (using for instance the United Kingdom Infrared Telescope and \textit{Spitzer}) and the dotted lines are BT-Settl models. For the L, T and Y dwarfs, the H$_2$O and CH$_4$ features are well marked. Figure from \citet{Allard2014}.}
	\label{bolmont_fig2}
	\end{figure}
	
BDs are very cold objects \citep[for example a 5~Gyr old BD of $0.02~\MSun$ has an effective temperature of $\sim 500$~K, see the grids of][]{Leconte2011} which means that their HZ is located very close in.
The cooling down of these objects also means that the HZ moves in with time, which has a strong importance for the potential habitability of planets.

Finally, their spectra are different from our Sun \citep[][]{Burrows2000, Allard2014}.
Due to their low temperature, their spectra are redshifted and display much more molecular lines than the Sun.
Fig.~\ref{bolmont_fig2} shows the spectra of different dwarfs (Spectral types M, L, T and Y).
The emission spectrum peaks around 1--2~$\mu$m for brown dwarfs, around 0.5~$\mu$m for the Sun and around $10~\mu$m for the Earth.
Decreasing the temperature and going from spectral type M to Y, we can see that the water and methane features become much more visible.
These spectral features have impacts on the potential climates of planets. 
For instance, the ice-albedo feedback, which is a positive feedback (lower temperature $\rightarrow$ more ice $\rightarrow$ higher albedo $\rightarrow$ less absorbed radiation $\rightarrow$ lower temperature) does not occur around red dwarfs due to the much lower value of the albedo of the ice in the IR \citep{JoshiHaberle2012}.

\section{Habitability of planets around brown dwarfs}

The inward migration of the HZ as the BD cools down has a major impact on potentially habitable planets: they might lose water early in their history.
Moreover, as BDs are cool objects, the HZ is close-in and planets are subjected to tides. 

\subsection{Before reaching the habitable zone}

\begin{figure}
	\begin{center}
	\includegraphics[width=9cm]{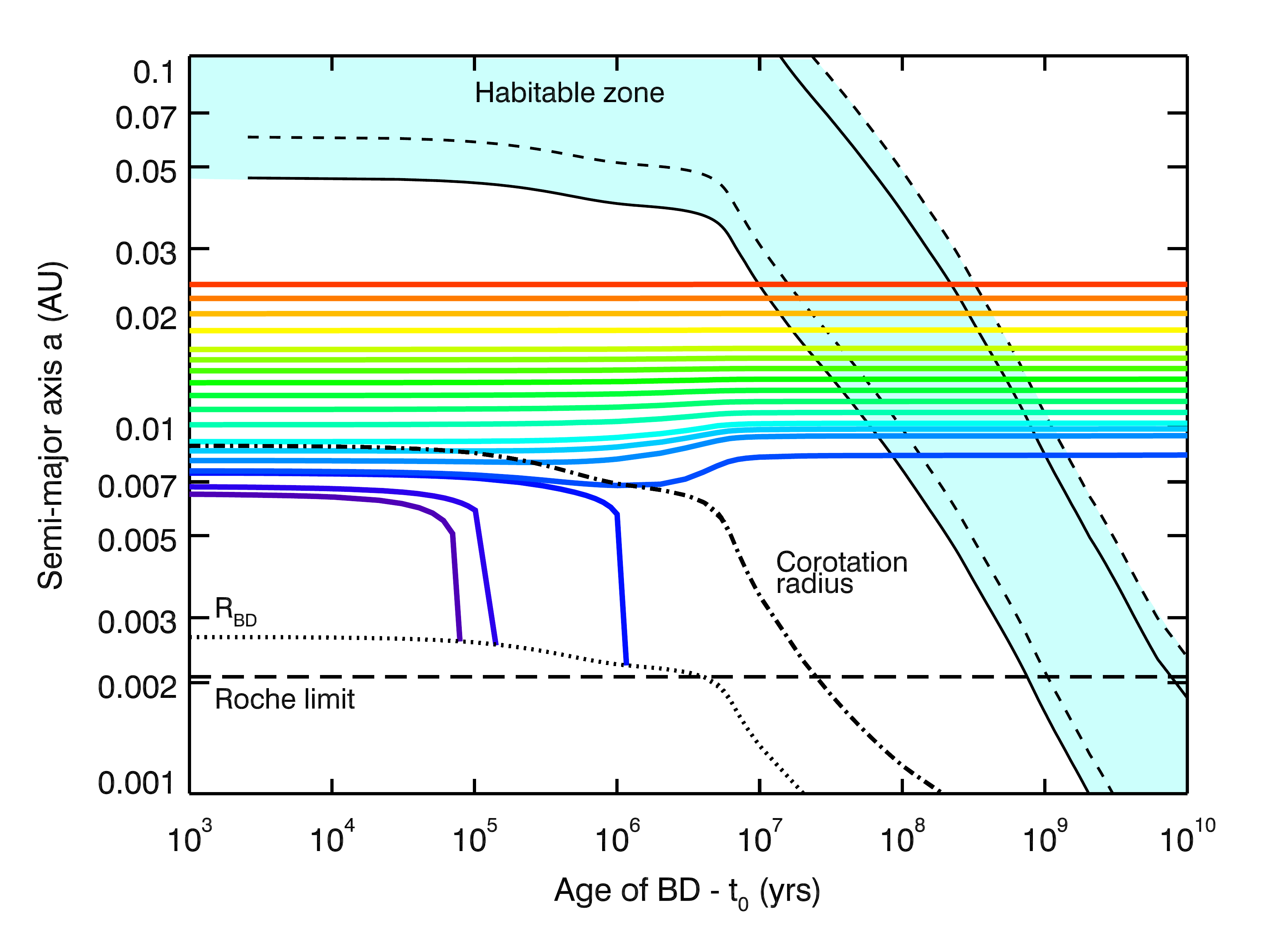}
	\end{center}
	\caption{Tidal evolution of Earth mass planets orbiting a $0.04~\MSun$ BD. The colored lines represent the orbital distance of different planets. The blue shaded area represents the HZ. The dashed-dotted line corresponds to the corotation radius. A planet farther than this limit migrates outward due to the tide it raises in the BD (just like the moon is migrating away from the Earth). The dotted line represents the radius of the BD. The long dashes represent the Roche limit: a planet closer than this limit would be tidally disrupted and create a ring of material around the BD (just like Saturn's rings, which are located inside its Roche limit).
	Figure adapted from \citet{Bolmont2011}.}
	\label{bolmont_fig3}
	\end{figure}

Planets in the HZ of Gyr-old BDs were initially too hot to host surface liquid water. 
Fig.~\ref{bolmont_fig3} shows the evolution of the HZ for a $0.04~\MSun$ BD.
The HZ moves inward with time so that planets spending some time in the HZ were all initially too hot for surface liquid water.
The closer-in planets spend a lot of time interior to the HZ (almost 100~Myr for the closest surviving planet in Fig.~\ref{bolmont_fig3}).
The farther the planet, the less time it spends interior to the HZ (10~Myr for the farthest planet in Fig.~\ref{bolmont_fig3}). 
During this time, all the water is in gaseous form in the atmosphere and is submitted to the high energy radiations from the BD. 
These radiations can break the water molecules and heat up the upper layers of the atmosphere to drive the escape of the hydrogen and oxygen atoms, which in the end results a net water loss.

The mechanisms driving the escape are complex and not well parametrized yet. 
However, a few studies \citep{BarnesHeller2013, LugerBarnes2015, Ribas2016, Bolmont2017} have tried to give an insight on these phenomenons to estimate the water loss from planets orbiting BDs and low-mass stars.
These studies differ on a few hypothesis but all use the variations of the same method which is based on the concept of the energy-limited escape mechanism \citep{Watson1981, Lammer2003}.
This mechanism relies on four steps: 1) the water molecules reach the upper regions of the atmosphere, 2) FUV radiation (100--200~nm) break the water molecules (photolysis), 3) XUV radiation (0.1--100~nm) heat up the upper layers of the atmosphere, 4) if the thermal velocity exceeds the escape velocity of the hydrogen and oxygen atoms, they can escape the planet. 
All these steps are considered to occur to compute the mass loss from the planets. 

The estimation of the FUV and XUV radiations are very observationally challenging for brown dwarfs.
Estimations of the X-ray luminosity exist for M-dwarfs \citep{Pizzolato2003} and for brown dwarfs \citep[e.g.][]{WilliamsCookBerger2014}, but the latter are actually mainly non-detections.
Besides, for some cool dwarfs, the Lyman-$\alpha$ emission ($121.6~$nm), which is a good proxy for the photolysis wavelength range, can be measured (see \citealt{Bourrier2017a} for TRAPPIST-1).
Very recently for TRAPPIST-1, the closest planet host we have to a brown dwarf, the following values have been obtained with the Space Telescope Imaging Spectrograph (STIS) on HST: $L_{\rm XUV} = 5.26-7.30 \times 10^{26}$~erg.s$^{-1}$ and $L_\alpha = 1.44-1.81\times 10^{26}$~erg.s$^{-1}$ \citep{Bourrier2017b}. 
The XUV luminosity is approximatively similar to that of Proxima Centauri but the Lyman-$\alpha$ emission is much lower \citep{Bourrier2017a, Bourrier2017b}.
TRAPPIST-1 might be at a transition between active M-dwarfs and more quiet brown dwarfs.

The emission of a M-dwarf and the emission of a brown dwarf have no reason to be similar, and taking into account their difference leads to somewhat different estimations for the water loss.
The studies about water loss from planets around low-mass stars and BDs mentioned before differ in the estimations of the high energy radiations.
While \citet{BarnesHeller2013} and \citet{LugerBarnes2015} considered XUV fluxes measured for early-type M-dwarfs \citep[e.g.][]{Pizzolato2003}, \citet{Ribas2016} and \citet{Bolmont2017} considered more recent estimations of the XUV fluxes for later-type M-dwarfs \citep[][]{WilliamsCookBerger2014, Osten2015}. 
Another main difference between these studies is that the latter used an estimation of the efficiency $\epsilon$ of the steps 3) and 4) based on 1D radiation-hydrodynamic mass-loss simulations \citep{OwenAlvarez2016}.

	\begin{figure}
	\begin{center}
	\includegraphics[width=\linewidth]{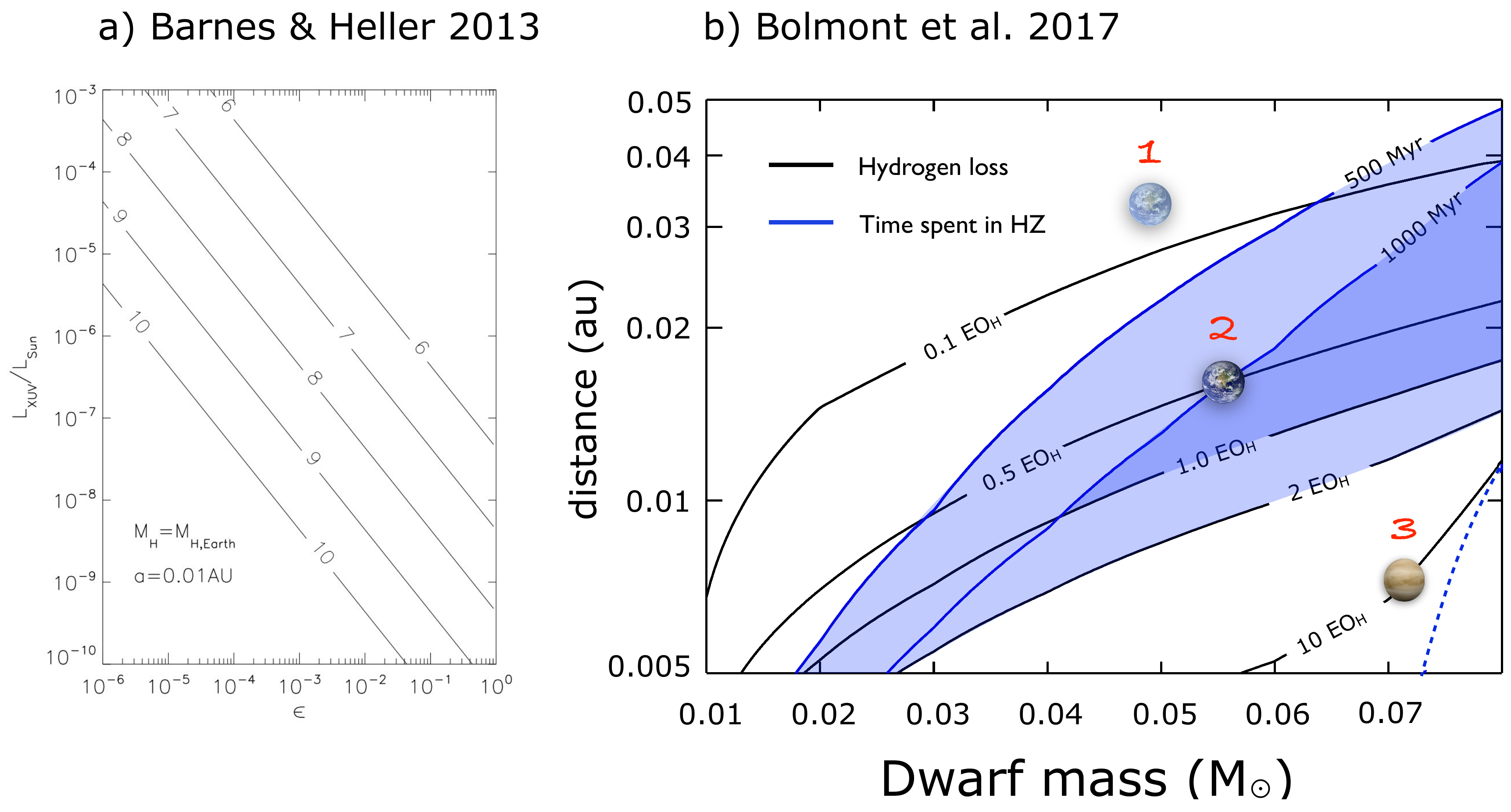}
	\end{center}
	\caption{a) Desiccation timescale for an Earth-like planet orbiting a BD at 0.01~au. Contour lines represent the logarithm of the time for the EarthÕs inventory of hydrogen to be lost (what we call here $1~EO_H$). $\epsilon$ is the efficiency of converting the XUV photons into the kinetic energy of escaping
particles. Figure from \citet{BarnesHeller2013}.
	b) Hydrogen loss from an Earth-like planet and time spent in the HZ (black and blue contours) for different masses of dwarfs and different planetary orbital distances. The blue shaded areas show the interesting regions of the parameter space: planets in this region lose little hydrogen (little water) before reaching the HZ and spend a long time in the HZ. Planet 1 loses a very small amount of water but spends less than 500~Myr in the HZ. Planet 2 loses a small amount of water and spends a lot time in the HZ. Planet 3 loses a lot of water and is probably desiccated once it reaches the HZ (unless its initial water reservoir was enormous). Figure adapted from \citet{Bolmont2017}}
	\label{bolmont_fig4}
	\end{figure}

Fig.~\ref{bolmont_fig4} shows the results obtained by a) \citet{BarnesHeller2013} and b) \citet{Bolmont2017}.
Due to the lack of observations and the uncertainty over the $\epsilon$ parameter (the efficiency of converting the XUV photons into the kinetic energy of escaping particles), \citet{BarnesHeller2013} explored the parameter space of the XUV luminosity (via $L_{\rm XUV}/L_{\rm bol}$) and $\epsilon$ (see Fig.~\ref{bolmont_fig4}a).
They concluded that the uncertainties are too big to rule in favor or against the presence of water on planets orbiting in the HZ of brown dwarfs.
In contrast, \citet{Bolmont2017} showed that there is a sweet spot for potentially habitable HZ planets: planets which do not lose a lot of water before reaching the HZ and which spend a long time in the HZ afterwards. 
Fig.~\ref{bolmont_fig4}b) shows the parameter space corresponding to this sweet spot.
When considering the range of possible XUV fluxes, the sweet spot moves around but is always existing and non negligible for the less massive planets.

To sum up, there is no consensus yet on the question of whether planets around BDs lose their water by the time they reach the HZ and the next step will likely come from observations.
Indeed, now that planets around very low-mass dwarfs are being discovered, the next tests will be to try to constrain their densities or try to detect water in the atmosphere of the very close-in planets. 
For instance, the masses and densities of the TRAPPIST-1 planets can be estimated with transit timing variations (\citealt{Gillon2017}, Grimm et al. submitted). 
A low density gives an indication on the presence of volatiles and the first estimates seem to be pointing in that direction for the TRAPPIST-1 planets.
The presence of water on these planets could be an indication that the water loss is overestimated in the studies done so far.
In such a context, the future observations of the JWST will be invaluable \citep{Barstow2016, Morley2017}.

\subsection{In the habitable zone}

Once the hot early phase has passed, one important factor for the eventual appearance of life is the time the planets actually spend inside the HZ. 
The most important parameters that influence the time a planet spends in the HZ are the orbital distance of the planet and the mass of the BD: the farther the planet the shorter the time in the HZ (see Figs.~\ref{bolmont_fig3} and \ref{bolmont_fig4}b) and the more massive the BD, the longer time in the HZ. 
However the orbital distance of the planet can evolve with time through the tidal interaction between the planet and the BD.

\medskip
\subsubsection{Time spent in the habitable zone vs tidal migration}

\begin{figure}
	\begin{center}
	\includegraphics[width=12cm]{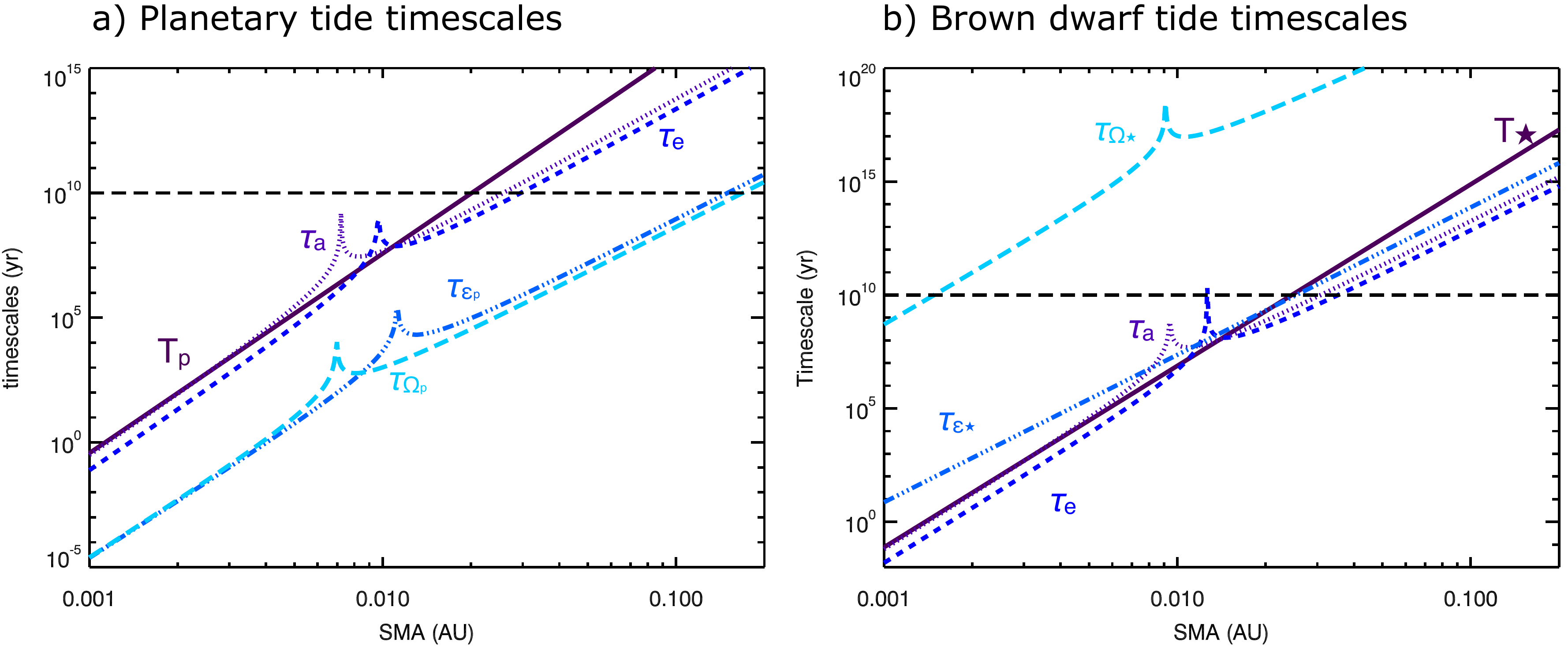}
	\end{center}
	\caption{Evolution timescales of the different quantities ($a$, $e$, $\Op$, $\epsilon_{\rm p}$, $\Os$, $\epsilon_\star$) impacted by tides for the a) planetary tide and b) brown dwarf tide. These timescales were calculated for an Earth-mass planet orbiting a $1$~Myr old $0.04~\MSun$ BD. The dissipation of the planet is equal to the Earth's \citep[same $\Delta t$, see][]{DeSurgyLaskar1997} and the dissipation of the BD is taken to be one of a hot Jupiter \citep{Hansen2010}. Figure from \citet{phd_Bolmont2013}.}
	\label{bolmont_fig5}
	\end{figure}

Tidal interactions are an important phenomenon that sculpts the architecture of close-in planetary systems. 
Both the tide raised by the planet in the BD (BD tide) and the tide raised by the BD in the planet (planetary tide) are playing a role in the evolution of the planetary system.  
Both tides influence the semi-major axis $a$ and eccentricity $e$ of the planet.
The planetary tide also influences the planet's rotation period $\Op$ and its obliquity $\epsilon_{\rm p}$ (the angle between the planet's rotation axis and the direction of the orbital angular momentum, the obliquity of the Earth is of about 23$^\circ$). 
The BD tide influences the inclination of the planet (or the BD obliquity $\epsilon_\star$) and the rotation $\Os$ of the BD.
Fig.~\ref{bolmont_fig5} shows the evolution timescales for the different quantities for a $1~\Mearth$ planet orbiting a $0.04~\MSun$ BD due to the a) planetary tide and b) BD tide.

The evolution timescales depend on the stellar and planetary parameters and the orbit parameters.
Among the parameters is the dissipation, which is a measure of how the system loses energy. 
This parameter is very poorly constrained and depends on the structural and rotational parameters of the body considered \citep[e.g. for stars,][]{Mathis2015}. 
Most tidal orbital studies use a simple model of equilibrium tide, where the dissipation is parametrized by a single parameter: a time lag $\Delta t$ in the constant time lag model \citep[e.g.,][]{Mignard1979, Hut1981, EKH1998}, or a quality factor $Q$ in the constant phase lag model \citep[e.g.,][]{GoldreichSoter1966}).
 
Fig.~\ref{bolmont_fig3} shows the tidal evolution of planets around a $0.04~\MSun$ BD. 
These planets are initially on circular orbits, with a zero obliquity and a synchronized rotation so only the BD tide influences their orbital evolution. 
The planets undergo an important tidal migration, which makes them either fall onto the BD or survive the early evolution and migrate outwards.
When the planets are in the HZ, their orbital distance is constant: the evolution timescale due to the BD tide has become so large (due to the small BD radius) that planets do not significantly migrate over timescales of several gigayears.

\begin{figure}
	\begin{center}
	\includegraphics[width=8cm]{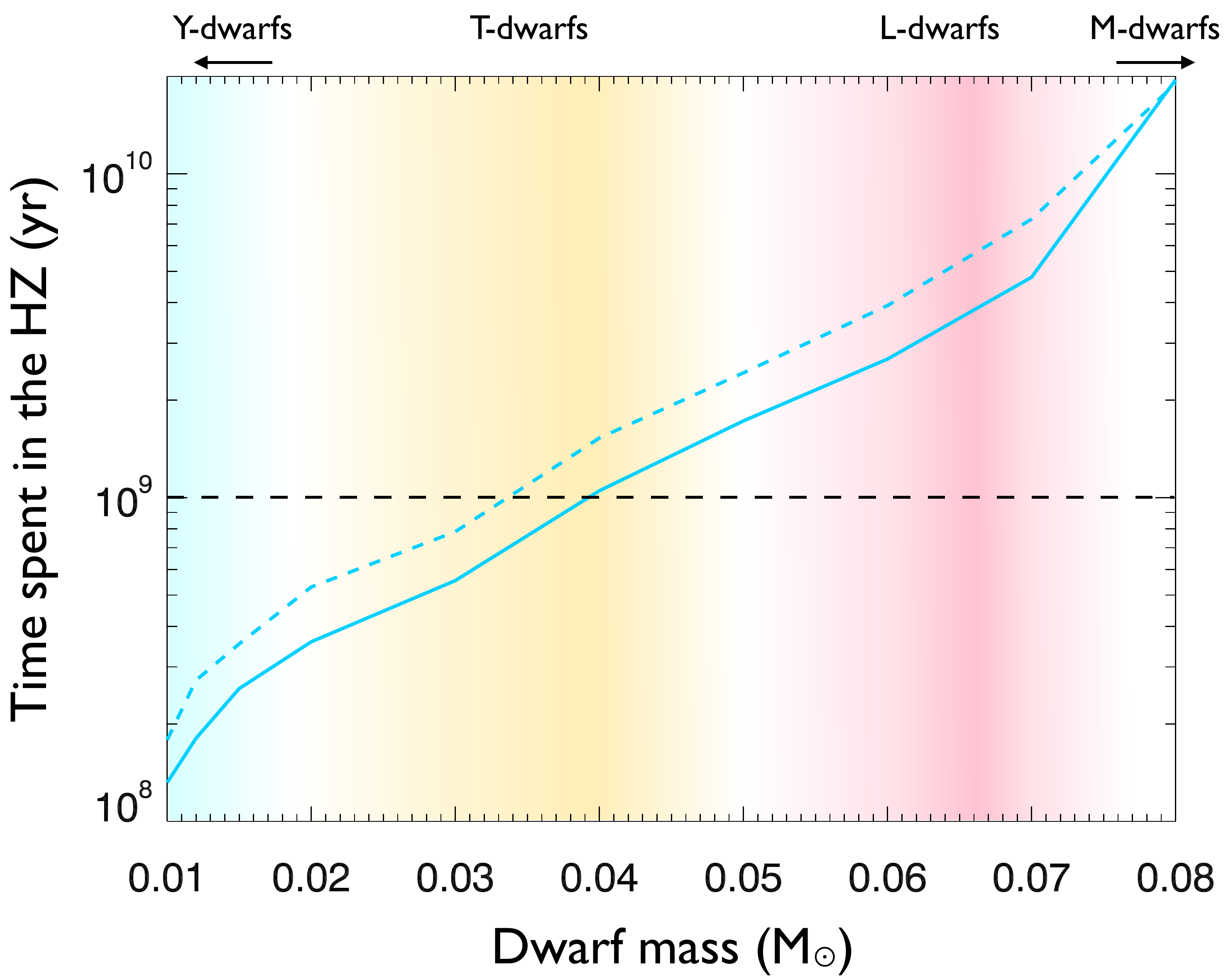}
	\end{center}
	\caption{Maximum time spent in the HZ for planets orbiting dwarfs of different masses (different spectral type). Figure adapted from \citet{Bolmont2011}.}
	\label{bolmont_fig6}
	\end{figure}

\citet{Andreeshchev2004} estimated the time a planet orbiting a BD can spend in the habitable zone. 
However, they did not take into account the tidal interactions between the BD and the planet.
\citet{Bolmont2011} and \citet{phd_Bolmont2013} showed that this interaction acts to decrease the time a planet can spend in the HZ.
They showed that 1) the higher the BD mass, 2) the higher the dissipation in the BD, 3) the higher the dissipation in the planet, the less time the planet spends in the HZ (See Fig.~\ref{bolmont_fig4}b for the effect of the mass of the BD).
However, despite that, they show that \textit{planets around BDs more massive than $0.04~\MSun$ could stay in the HZ up to a few gigayears} (see Figure~\ref{bolmont_fig6}), leaving plenty of time for life to potentially appear and evolve \citep{Bolmont2011}.

\subsubsection{One planet system vs multiple planet system}

However, once the planets reach the HZ and assuming they could retain a sufficient part of their initial water reservoir, the presence of surface liquid water is still not yet assured.

Let us first consider that there is only one planet in the system.
Fig.~\ref{bolmont_fig5} shows that by the time close-in planets reach the HZ (after a few 10~Myr to 100~Myr, see Fig.~\ref{bolmont_fig3}), planetary tides have had time to damp the initial obliquity, synchronize the rotation and damp the eccentricity so that planets are on a circular orbit, have a zero obliquity and are tidally locked.
The planet therefore always shows the same side to the BD and its poles receive very little light. 
This raises the problem of the possible existence of regions on the planet where the temperature is constantly lower than the melting point of water and where all the water of the planet will condense \citep[the so called ``cold traps'', e.g.][]{Joshi2003}. 
In this configuration, the night side and the poles could be cold traps and the planet would not be able to host surface liquid water. 

However, there are mechanisms which can prevent the appearance of cold traps or that can prevent synchronization altogether. 
For instance, the existence of a dense enough atmosphere would allow a better heat repartition and allow surface liquid water just as \citet{Wordsworth2011} showed for Gliese~581~d (a super-Earth orbiting a $0.3~\MSun$ dwarf). 
Also, if planets are synchronized on a close-in orbit, their rotation could trigger winds that would efficiently redistribute heat to the night side \citep{Showman2011, Leconte2013, Bolmont2016}.
Recently, \citet{Turbet2016} and \citet{Turbet2017} respectively showed that Proxima-b and TRAPPIST-1 e, f, g could have surface liquid water despite a synchronous rotation.
Furthermore, a high enough geothermal flux could also prevent the appearance of cold traps.
Finally, recent works \citep[e.g.][]{Leconte2015} showed that the atmospheric tides could act to desynchronize the rotation for planets around low-mass stars.
Atmospheric tides are different from gravitational tides: the repartition of mass in the atmosphere is due to the irradiation from the star not the gravitational pull of the star \citep[e.g.][]{GoldSoter1969}.
\citet{Auclair-Desrotour2017} showed that the outcome of atmospheric tides actually depends the stability of atmospheric layers close to the ground (only a convective atmosphere can act to desynchronize the rotation).
The prediction of the rotation state of HZ planets is therefore not straightforward. 
JWST observations could help establish the presence of an atmosphere and distinguish between a convective atmosphere or a stably stratified atmosphere, which would tell us if the planet is likely to be synchronized or not.  
Note that \citet{Leconte2015} also showed that for stars of mass lower than $0.3~\MSun$, gravitational tides might prevail on atmospheric tides, but this should be investigated further.

\medskip

The situation differs significantly if the planet is part of a multiple planet system. 
Due to planet-planet interactions, both eccentricity and obliquity do not tend to 0 but to an equilibrium value which is the result of the competition between planet-planet excitation and tidal damping. 

An extreme case of planet-planet interactions is the Mean Motion Resonance (MMR), which happens when the ratio between the orbital period of two planets is commensurable. 
One of the consequences is that the eccentricity of both planets is excited to higher levels. 
A close-by example is the 1:2:4 MMR between Io, Europa and Ganymede. 
The resonance maintains a non-negligible eccentricity and causes Io to experience an intense internal heating due to the stress it experiences on one orbit. 
\citet{Spencer2000} estimated the internal heat flux of Io to be around $3$~W/m$^2$. 
This high heat flux is responsable for the intense volcanic activity \citep[e.g. the Tvashtar volcano,][]{Spencer2007}.
To give a point of comparison, the internal heat flux of Earth is about 40 times lower than Io \citep[about $0.08$~W/m$^2$ but mainly due to radioactivity; e.g.,][]{Davies2010}. 
A tidally evolving planet in the HZ of a BD could thus experience such an intense tidal heating that it can 
have repercussions on the internal structure of the planet (Mantle overheating, e.g. \citealt{Behounkova2011} and \citealt{HenningHurford2014}; Effect on the planetary magnetic field, e.g. \citealt{DriscollBarnes2015}) and it can drive the atmosphere of the planet in a runaway greenhouse state.
\citet{Jackson2008}, \citet{Barnes2009, Barnes2013} investigated this latter phenomenon for planets around M-dwarfs and more massive stars and introduced the notion of ``tidal habitable zone'' and ``tidal Venus'' planets: they are in the classical HZ but have a tidal heat flux higher than 309~W~m$^{-2}$, which triggers the runaway greenhouse state.

\begin{figure}
	\begin{center}
	\includegraphics[width=10cm]{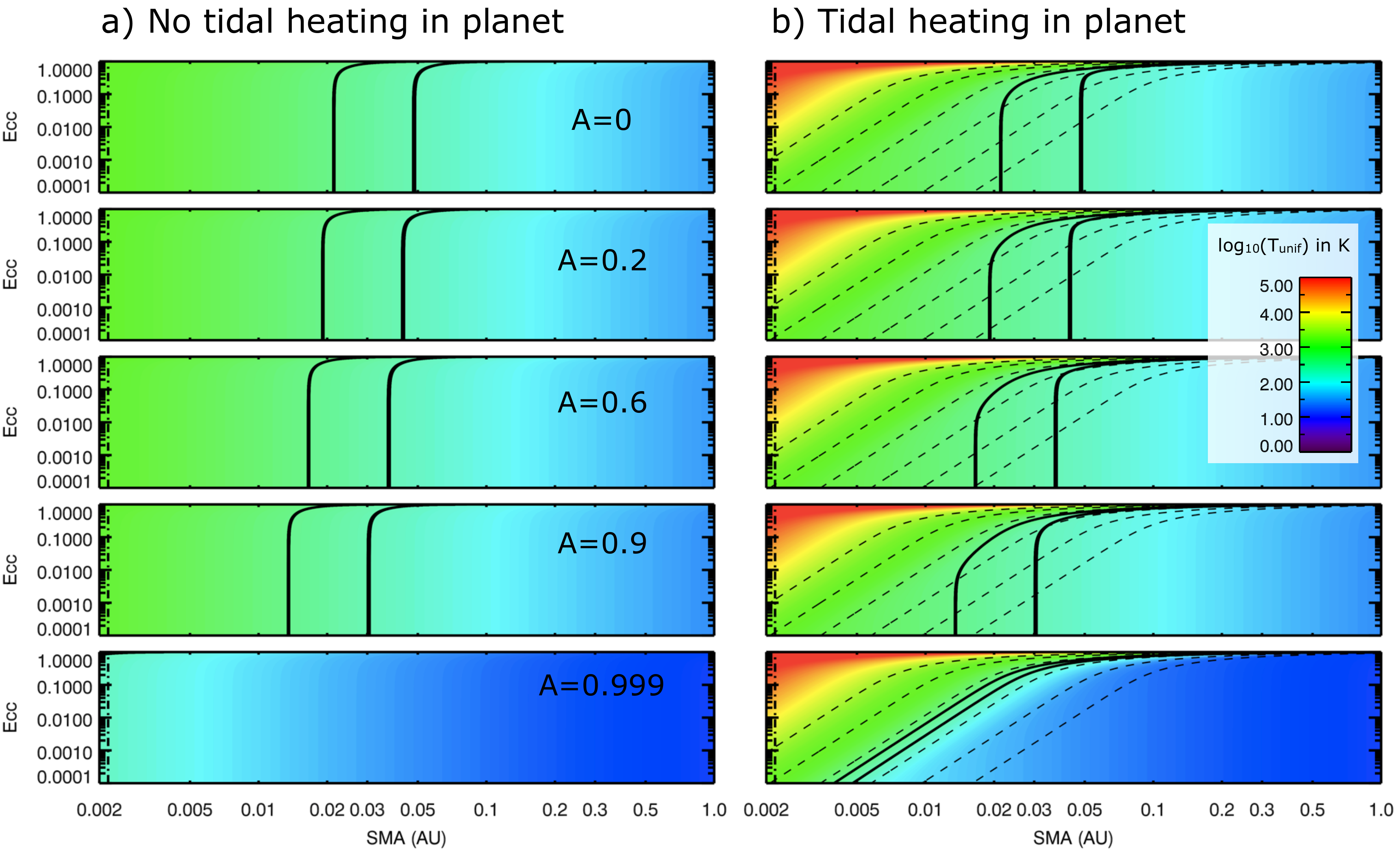}
	\end{center}
	\caption{Maps of the log$_{10}$ of the temperature of a $1~\Mearth$ planet orbiting a $100$~Myr old $0.04~\MSun$ BD as a function of its orbital distance and the eccentricity of its orbit. a) The tidal heating is not taken into account. b) The tidal heating is taken into account. The different panels are for a different albedo of the planet (A increases from top to bottom). The two black lines correspond to temperatures of 180~K (flux of 240~W.m$^{-2}$) and 270~K (flux of 300~W.m$^{-2}$), which crudely represent the limits of the HZ. Figure adapted from \citet{phd_Bolmont2013}.}
	\label{bolmont_fig7}
	\end{figure}

\citet{phd_Bolmont2013} investigated the effect of tides on the HZ limit around BDs for different eccentricities and different albedos for the planet.
Fig.~\ref{bolmont_fig7} shows the HZ limit for a planet orbiting a $0.04~\MSun$ BD a) not taking into account tidal heating b) taking into account tidal heating \citep[and assuming a dissipation equal to the Earth's,][]{DeSurgyLaskar1997}. 
Eccentricity has an effect on the HZ limits: the higher the eccentricity, the farther the HZ limits (see Fig.~\ref{bolmont_fig7}a).
When taking into account tidal heating the HZ inner limit is strongly impacted. 
Overall, tidal heating has the effect of narrowing the HZ by pushing away the inner edger more than the outer edge.
While \citet{phd_Bolmont2013} did not investigate the effect of the obliquity, note that a non-zero obliquity would also act as to push and narrow the HZ even more.  

	\begin{figure}
	\centering
	\includegraphics[width=0.9\textwidth,clip]{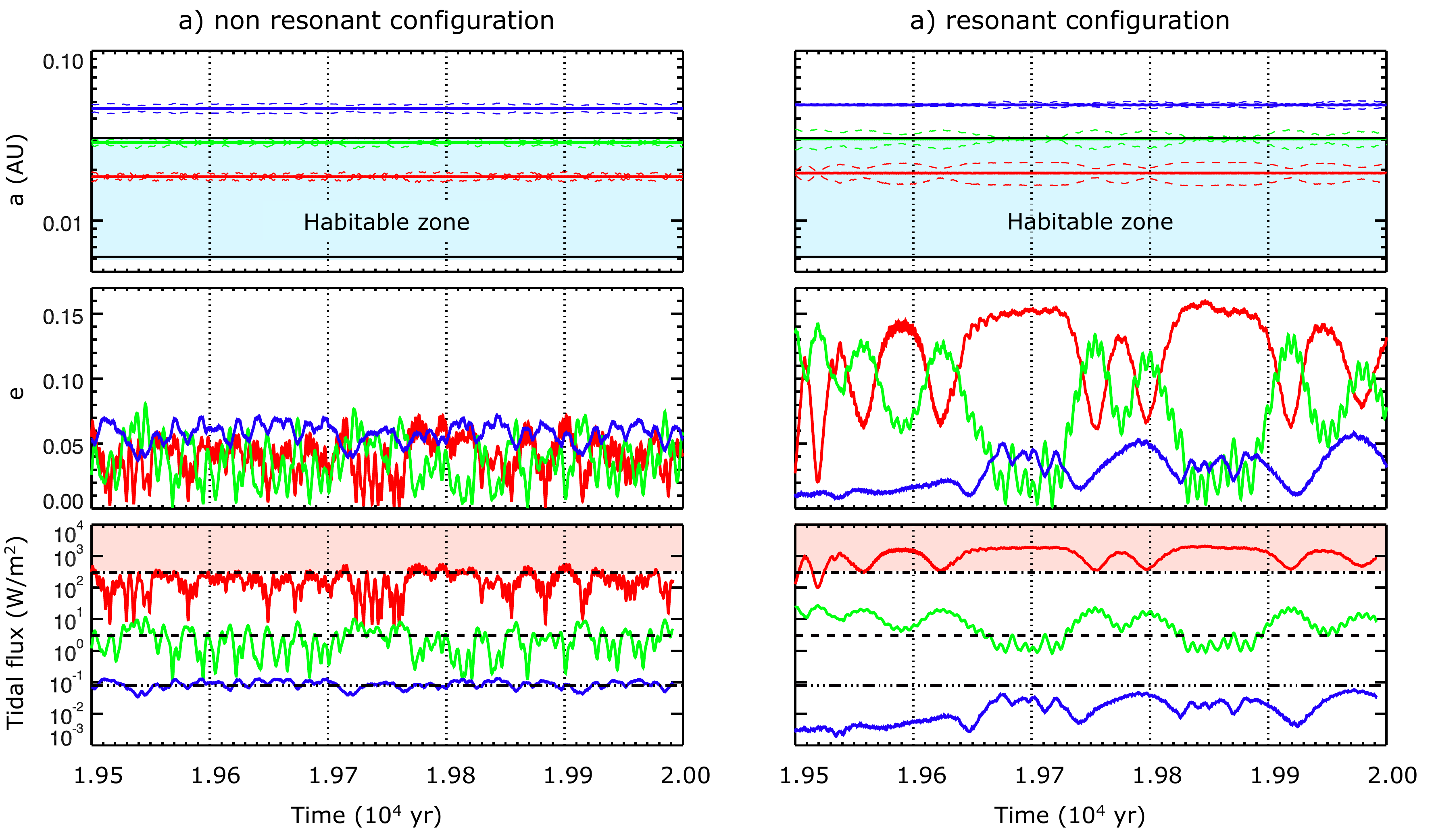}      
	\caption{Evolution of the orbital distance, eccentricity and tidal heat flux of three Earth-sized planets orbiting a BD of $0.08~M_\star$ in a) a non-resonant configuration, b) a resonant configuration (1:2 MMR). Top graph: the full colored lines correspond to the semi-major axis evolution of the 3 planets, and the dashed lines correspond to their perihelion and aphelion distances. The blue shaded region is the HZ. Middle graph: eccentricity of the 3 planets. Bottom graph: the full colored lines correspond to the tidal heat flux of the 3 planets. The black dashed dotted line corresponds to the limit of runaway greenhouse \citep[e.g.][]{Kopparapu2013}, the dashed line corresponds to Io's heat flux and the dashed 3 dots line corresponds to Earth's heat flux. The shaded red region corresponds to where the heat flux is so high that the planet is in a runaway greenhouse state. Figure from \citet{Bolmont2014sf2a}.}
	\label{bolmont_fig8}
	\end{figure}

\medskip

In multiple planet systems, the eccentricity and obliquity can be maintained to high enough levels so that tidal heating has an impact on the potential of the planet to host surface liquid water.
\citet{Bolmont2014sf2a} extended the works of \citet{Barnes2013} to treat the specific case of multiple planet systems around BD.
They illustrated the importance of tides by considering the case of a system of three Earth-sized planets orbiting just outside the corotation radius of a $0.08~M_\star$ dwarf for two different tidal dissipation factors.
The planets experience a convergent outward migration, which leads either to a resonant capture for a high BD dissipation or not for low dissipation.
In the case of a high BD dissipation, they found that the planets enter a MMR chain (1:2:4) in a few million years of evolution.

Fig.~\ref{bolmont_fig8} shows the short term evolution of such a system at an age of 1~Gyr for the two cases. 
The two inner planets are in the HZ. 
The eccentricities of the planets in case a) are relatively small $<0.07$ but in case b) due to the MMR excitation, they can reach $0.15$. 
In case a), the average of the tidal heat flux of the inner planet remains below the runaway greenhouse limit.
In case b), its tidal heat flux is almost always above the runaway greenhouse limit: in spite of being in the HZ, it would therefore be a ``tidal Venus'' and would be too hot to host surface liquid water. 
The middle planet (in green), also in the HZ, with a flux higher than Io's in case b) would experience a intense volcanism, which could be problematic for potential life.
Conversely, this planet spends some time around aphelion outside the insolation HZ and could be too cold to be able to sustain a potential liquid water reservoir. 
However taking into account tidal heating could improve the conditions for habitability at apocenter. 
One could imagine a more extreme case of a planet on an orbit completely outside the HZ but heated up by tides sufficiently to be able to host surface liquid water.

This mechanism can facilitate the habitable conditions for planets on the outer edge of the HZ or even exterior to the HZ. 
Recently, Ramirez \& Kaltenegger 2017 showed that volcanoes ejecting hydrogen in the atmosphere in a regular way could contribute to extend the HZ farther than the classical limits. 
Such volcanism maintained by tides in a multi-planet system could therefore be favorable to surface liquid water conditions in the colder regions of a system.

Therefore, when assessing habitability of planets in the HZ of BDs, one should investigate if tides are strong enough to drive a runaway greenhouse \citep{Barnes2013, Bolmont2014sf2a}. 
If the planet absorbs an average flux ($\Phi_\star + \Phi_{{\rm tides}}$) lower than the greenhouse limit, the planet can sustain a liquid water reservoir, but if it receives an average flux higher than the greenhouse limit the planet will be too hot to be able to sustain a liquid water reservoir.

\subsection{Brown dwarfs' variability}

Some brown dwarfs are known to be variable objects at various wavelengths: from near-IR \citep[e.g.][thought to be due to clouds or spots]{Artigau2009} to X-ray \citep{Rutledge2000}.
While the photometric variability is of relatively low amplitude \citep{Buenzli2014} and probably does not have a significant impact on the atmosphere, the energetic flares can potentially have a negative effect on the atmosphere (driving water loss as during the runaway greenhouse phase, see previous section) and life (e.g. \citealt{Tabataba-Vakili2016} for M-dwarfs).
 
The effect of energetic flares has been widely discussed for M-dwarfs. 
For instance, UV flares can lead to ozone depletion, which increases the penetration of the UV photons and can damage eventual surface life \citep{Segura2010}. 
Depending on their frequency, the flares can also alter the chemistry of the planet preventing it from reaching an equilibrium \citep{Venot2016, Segura2010}.
Recently, \citet{Vida2017} and \citet{O'Malley-James2017} estimated that the UV environment of the TRAPPIST-1 system might be too harsh for life.

However, brown dwarfs might be quieter than M-dwarfs \citep{Mohanty2002, WilliamsCookBerger2014}.
Besides, the measured activity in itself might not be representative of what the planet actually receives during flaring events: the planet is impacted only if the flare is in its direction \citep{Segura2010}.
Furthermore, the effect of these energetic radiations also strongly depends on the atmosphere of the planet \citep[e.g.][]{O'Malley-James2017} and its magnetic field \citep[e.g.][]{Kay2016}.
Finally, despite all the drawbacks of receiving energetic radiations, \citet{Ranjan2017} also pointed out that some UV might be required for the origin of life.

The questions of the influence of harmful flares on the planetary environment of planets around cool dwarfs will benefit both from the future observations of the atmospheric chemistry of these objects and the necessary constraints on the dwarf's complete spectrum.


\section{Observational perspectives for planets around brown dwarfs (and more generally cool dwarfs)}

The future prospects of observation and characterization of planets around BD have never been better \citep{He2017}. 
A few missions are either dedicated to planets orbiting very faint objects: TRAPPIST \citep{Gillon2011},  SPECULOOS \citep{Gillon2013} and SPIRou \citep{Artigau2011}; or able to observe them such as the K2 mission \citep{Haas2014} and Spitzer \citep[as proposed by][based on a study of \citealt{Belu2013}]{Triaud2013}.

The recent discovery of the multiple planet system around TRAPPIST-1 \citep{Gillon2016, Gillon2017} highlights the importance of studying these objects. 
TRAPPIST-1 is a quasi BD, just above the theoretical limit between BDs and M-dwarfs, and illustrates the fact that planets will probably be found around BDs in the near future \citep[e.g., with SPECULOOS,][]{Gillon2013}.
Planetary systems around very low mass stars and BDs (hereafter ultra-cool dwarfs) are dynamically rich: the planets are tidally evolving, most systems are compact and therefore planet-planet interactions play a major role.  

What makes the planetary systems orbiting ultra-cool dwarfs even more interesting is the prospect of future observations.
Indeed the planets in the HZ of cool dwarfs are the only HZ planets whose atmosphere can be probed by telescopes such as the JWST \citep{Belu2013}.
For instance, \citet{Barstow2016} recently showed that ozone could be detected in the atmosphere of the three inner planets of TRAPPIST-1 with a high number of transits (at least 60 for TRAPPIST-1c).
\citet{Morley2017} also showed that it could be potentially possible to differentiate between an Earth-like, a Venus-like and a Titan-like atmosphere with JWST observations of TRAPPIST-1c.
For non-transiting planets, there are also possibilities via emission spectroscopy with the E-ELT and emission phase curves with the JWST \citep[see][for a discussion about Proxima-b]{Turbet2016}.

The era of planets around BDs is almost upon us and these objects will represent a highly interesting scientific domain.
Rocky planets around BDs (and very low-mass stars, such as the TRAPPIST-1 planetary system), will allow us to do comparative planetology and explore in a unique way the effect of tidal orbital dynamics on the potential climate of these planets.

\begin{acknowledgement}
E.B. acknowledges funding by the European Research Council through ERC grant SPIRE 647383. 
This research has made use of NASA's Astrophysics Data System.
\end{acknowledgement}


\begin{thebibliography}{111}
\providecommand{\natexlab}[1]{#1}
\providecommand{\url}[1]{{#1}}
\providecommand{\urlprefix}{URL }
\expandafter\ifx\csname urlstyle\endcsname\relax
  \providecommand{\doi}[1]{DOI~\discretionary{}{}{}#1}\else
  \providecommand{\doi}{DOI~\discretionary{}{}{}\begingroup
  \urlstyle{rm}\Url}\fi
\providecommand{\eprint}[2][]{\url{#2}}

\bibitem[{{Allard}(2014)}]{Allard2014}
{Allard} F (2014) {The BT-Settl Model Atmospheres for Stars, Brown Dwarfs and
  Planets}. In: {Booth} M, {Matthews} BC {Graham} JR (eds) Exploring the
  Formation and Evolution of Planetary Systems, IAU Symposium, vol 299, pp
  271--272, \doi{10.1017/S1743921313008545}

\bibitem[{{Andreeshchev} and {Scalo}(2004)}]{Andreeshchev2004}
{Andreeshchev} A {Scalo} J (2004) {Habitability of Brown Dwarf Planets}. In:
  {R~Norris \& F~Stootman} (ed) Bioastronomy 2002: Life Among the Stars, IAU
  Symposium, vol 213, pp 115--+

\bibitem[{{Anglada-Escud{\'e}} et~al.(2016){Anglada-Escud{\'e}}, {Amado},
  {Barnes}, {Berdi{\~n}as}, {Butler}, {Coleman}, {de La Cueva}, {Dreizler},
  {Endl}, {Giesers}, {Jeffers}, {Jenkins}, {Jones}, {Kiraga}, {K{\"u}rster},
  {L{\'o}pez-Gonz{\'a}lez}, {Marvin}, {Morales}, {Morin}, {Nelson}, {Ortiz},
  {Ofir}, {Paardekooper}, {Reiners}, {Rodr{\'{\i}}guez},
  {Rodr{\'{\i}}guez-L{\'o}pez}, {Sarmiento}, {Strachan}, {Tsapras}, {Tuomi},
  and {Zechmeister}}]{Anglada-Escude2016}
{Anglada-Escud{\'e}} G, {Amado} PJ, {Barnes} J et~al. (2016) {A terrestrial
  planet candidate in a temperate orbit around Proxima Centauri}. \nat
  536:437--440

\bibitem[{{Artigau} et~al.(2009){Artigau}, {Bouchard}, {Doyon}, and
  {Lafreni{\`e}re}}]{Artigau2009}
{Artigau} {\'E}, {Bouchard} S, {Doyon} R {Lafreni{\`e}re} D (2009) {Photometric
  Variability of the T2.5 Brown Dwarf SIMP J013656.5+093347: Evidence for
  Evolving Weather Patterns}. \apj 701:1534--1539

\bibitem[{{Artigau} et~al.(2011){Artigau}, {Donati}, and
  {Delfosse}}]{Artigau2011}
{Artigau} {\'E}, {Donati} JF {Delfosse} X (2011) {Planet Detection, Magnetic
  Field of Protostars and Brown Dwarfs Meteorology with SPIRou}. In:
  {Johns-Krull} C, {Browning} MK {West} AA (eds) 16th Cambridge Workshop on
  Cool Stars, Stellar Systems, and the Sun, Astronomical Society of the Pacific
  Conference Series, vol 448, p 771

\bibitem[{{Auclair-Desrotour} et~al.(2017){Auclair-Desrotour}, {Laskar}, and
  {Mathis}}]{Auclair-Desrotour2017}
{Auclair-Desrotour} P, {Laskar} J {Mathis} S (2017) {Atmospheric tides in
  Earth-like planets}. \aap 603:A107

\bibitem[{{Barnes} and {Heller}(2013)}]{BarnesHeller2013}
{Barnes} R {Heller} R (2013) {Habitable Planets Around White and Brown Dwarfs:
  The Perils of a Cooling Primary}. Astrobiology 13:279--291

\bibitem[{{Barnes} et~al.(2008){Barnes}, {Raymond}, {Jackson}, and
  {Greenberg}}]{Barnes2008}
{Barnes} R, {Raymond} SN, {Jackson} B {Greenberg} R (2008) {Tides and the
  Evolution of Planetary Habitability}. Astrobiology 8:557--568

\bibitem[{{Barnes} et~al.(2009){Barnes}, {Jackson}, {Greenberg}, and
  {Raymond}}]{Barnes2009}
{Barnes} R, {Jackson} B, {Greenberg} R {Raymond} SN (2009) {Tidal Limits to
  Planetary Habitability}. \apjl 700:L30--L33

\bibitem[{{Barnes} et~al.(2010){Barnes}, {Jackson}, {Greenberg}, {Raymond}, and
  {Heller}}]{Barnes2010ASPC}
{Barnes} R, {Jackson} B, {Greenberg} R, {Raymond} SN {Heller} R (2010) {Tidal
  Constraints on Planetary Habitability}. In: {Coud{\'e} du Foresto} V,
  {Gelino} DM {Ribas} I (eds) Pathways Towards Habitable Planets, Astronomical
  Society of the Pacific Conference Series, vol 430, p 133

\bibitem[{{Barnes} et~al.(2011){Barnes}, {Meadows}, {Domagal-Goldman},
  {Heller}, {Jackson}, {L{\'o}pez-Morales}, {Tanner}, {G{\'o}mez-P{\'e}rez},
  and {Ruedas}}]{Barnes2011ASPC}
{Barnes} R, {Meadows} VS, {Domagal-Goldman} SD et~al. (2011) {Habitability of
  Planets Orbiting Cool Stars}. In: {Johns-Krull} C, {Browning} MK {West} AA
  (eds) 16th Cambridge Workshop on Cool Stars, Stellar Systems, and the Sun,
  Astronomical Society of the Pacific Conference Series, vol 448, p 391

\bibitem[{{Barnes} et~al.(2013){Barnes}, {Mullins}, {Goldblatt}, {Meadows},
  {Kasting}, and {Heller}}]{Barnes2013}
{Barnes} R, {Mullins} K, {Goldblatt} C et~al. (2013) {Tidal Venuses: Triggering
  a Climate Catastrophe via Tidal Heating}. Astrobiology 13:225--250

\bibitem[{{Barrado y Navascu{\'e}s} et~al.(2001){Barrado y Navascu{\'e}s},
  {Stauffer}, {Brice{\~n}o}, {Patten}, {Hambly}, and {Adams}}]{Barrado2001}
{Barrado y Navascu{\'e}s} D, {Stauffer} JR, {Brice{\~n}o} C et~al. (2001) {Very
  Low-Mass Stars and Brown Dwarfs of the Young Open Cluster IC 2391}. ApJS
  134:103--114

\bibitem[{{Barstow} and {Irwin}(2016)}]{Barstow2016}
{Barstow} JK {Irwin} PGJ (2016) {Habitable worlds with JWST: transit
  spectroscopy of the TRAPPIST-1 system?} \mnras 461:L92--L96

\bibitem[{{Belu} et~al.(2013){Belu}, {Selsis}, {Raymond}, {Pall{\'e}},
  {Street}, {Sahu}, {von Braun}, {Bolmont}, {Figueira}, {Anupama}, and
  {Ribas}}]{Belu2013}
{Belu} AR, {Selsis} F, {Raymond} SN et~al. (2013) {Habitable Planets Eclipsing
  Brown Dwarfs: Strategies for Detection and Characterization}. ApJ 768:125

\bibitem[{Bolmont(2013)}]{phd_Bolmont2013}
Bolmont E (2013) Evolution et habitabilit\'e de syst\`emes plan\'etaires autour
  d'\'etoiles de faible masse et de naines brunes. PhD thesis, Universit\'e de
  Bordeaux 1, Universit\'e de Bordeaux 1, Pessac, an optional note

\bibitem[{{Bolmont} et~al.(2011){Bolmont}, {Raymond}, and
  {Leconte}}]{Bolmont2011}
{Bolmont} E, {Raymond} SN {Leconte} J (2011) {Tidal evolution of planets around
  brown dwarfs}. A \& A 535:A94

\bibitem[{{Bolmont} et~al.(2014){Bolmont}, {Raymond}, and
  {Selsis}}]{Bolmont2014sf2a}
{Bolmont} E, {Raymond} SN {Selsis} F (2014) {Dynamics of exoplanetary systems,
  links to their habitability}. In: {Ballet} J, {Martins} F, {Bournaud} F,
  {Monier} R {Reyl{\'e}} C (eds) SF2A-2014: Proceedings of the Annual meeting
  of the French Society of Astronomy and Astrophysics, pp 63--68

\bibitem[{{Bolmont} et~al.(2016){Bolmont}, {Libert}, {Leconte}, and
  {Selsis}}]{Bolmont2016}
{Bolmont} E, {Libert} AS, {Leconte} J {Selsis} F (2016) {Habitability of
  planets on eccentric orbits: Limits of the mean flux approximation}. \aap
  591:A106

\bibitem[{{Bolmont} et~al.(2017){Bolmont}, {Selsis}, {Owen}, {Ribas},
  {Raymond}, {Leconte}, and {Gillon}}]{Bolmont2017}
{Bolmont} E, {Selsis} F, {Owen} JE et~al. (2017) {Water loss from terrestrial
  planets orbiting ultracool dwarfs: implications for the planets of
  TRAPPIST-1}. \mnras 464:3728--3741

\bibitem[{{Bonfils} et~al.(2013){Bonfils}, {Delfosse}, {Udry}, {Forveille},
  {Mayor}, {Perrier}, {Bouchy}, {Gillon}, {Lovis}, {Pepe}, {Queloz}, {Santos},
  {S{\'e}gransan}, and {Bertaux}}]{Bonfils2013}
{Bonfils} X, {Delfosse} X, {Udry} S et~al. (2013) {The HARPS search for
  southern extra-solar planets. XXXI. The M-dwarf sample}. A\&A 549:A109

\bibitem[{{Boss} et~al.(2007){Boss}, {Butler}, {Hubbard}, {Ianna},
  {K{\"u}rster}, {Lissauer}, {Mayor}, {Meech}, {Mignard}, {Penny},
  {Quirrenbach}, {Tarter}, and {Vidal-Madjar}}]{Boss2007}
{Boss} AP, {Butler} RP, {Hubbard} WB et~al. (2007) {Working Group on Extrasolar
  Planets}. Transactions of the International Astronomical Union, Series A
  26:183--186

\bibitem[{{Bourrier} et~al.(2017{\natexlab{a}}){Bourrier}, {de Wit}, {Bolmont},
  {Stamenkovi{\'c}}, {Wheatley}, {Burgasser}, {Delrez}, {Demory}, {Ehrenreich},
  {Gillon}, {Jehin}, {Leconte}, {Lederer}, {Lewis}, {Triaud}, and {Van
  Grootel}}]{Bourrier2017b}
{Bourrier} V, {de Wit} J, {Bolmont} E et~al. (2017{\natexlab{a}}) {Temporal
  Evolution of the High-energy Irradiation and Water Content of TRAPPIST-1
  Exoplanets}. \aj 154:121

\bibitem[{{Bourrier} et~al.(2017{\natexlab{b}}){Bourrier}, {Ehrenreich},
  {Wheatley}, {Bolmont}, {Gillon}, {de Wit}, {Burgasser}, {Jehin}, {Queloz},
  and {Triaud}}]{Bourrier2017a}
{Bourrier} V, {Ehrenreich} D, {Wheatley} PJ et~al. (2017{\natexlab{b}})
  {Reconnaissance of the TRAPPIST-1 exoplanet system in the Lyman-{$\alpha$}
  line}. \aap 599:L3

\bibitem[{{Buenzli} et~al.(2014){Buenzli}, {Apai}, {Radigan}, {Reid}, and
  {Flateau}}]{Buenzli2014}
{Buenzli} E, {Apai} D, {Radigan} J, {Reid} IN {Flateau} D (2014) {Brown Dwarf
  Photospheres are Patchy: A Hubble Space Telescope Near-infrared Spectroscopic
  Survey Finds Frequent Low-level Variability}. \apj 782:77

\bibitem[{{Burrows} et~al.(2000){Burrows}, {Marley}, and {Sharp}}]{Burrows2000}
{Burrows} A, {Marley} MS {Sharp} CM (2000) {The Near-Infrared and Optical
  Spectra of Methane Dwarfs and Brown Dwarfs}. \apj 531:438--446

\bibitem[{{B{\v e}hounkov{\'a}} et~al.(2011){B{\v e}hounkov{\'a}}, {Tobie},
  {Choblet}, and {{\v C}adek}}]{Behounkova2011}
{B{\v e}hounkov{\'a}} M, {Tobie} G, {Choblet} G {{\v C}adek} O (2011) {Tidally
  Induced Thermal Runaways on Extrasolar Earths: Impact on Habitability}. ApJ
  728:89

\bibitem[{{Caballero} et~al.(2007){Caballero}, {B{\'e}jar}, {Rebolo},
  {Eisl{\"o}ffel}, {Zapatero Osorio}, {Mundt}, {Barrado Y Navascu{\'e}s},
  {Bihain}, {Bailer-Jones}, {Forveille}, and {Mart{\'{\i}}n}}]{Caballero2007}
{Caballero} JA, {B{\'e}jar} VJS, {Rebolo} R et~al. (2007) {The substellar mass
  function in {$\sigma$} Orionis. II. Optical, near-infrared and IRAC/Spitzer
  photometry of young cluster brown dwarfs and planetary-mass objects}. A \& A
  470:903--918

\bibitem[{{Chabrier}(2002)}]{Chabrier2002}
{Chabrier} G (2002) {The Galactic Disk Mass Budget. II. Brown Dwarf Mass
  Function and Density}. \apj 567:304--313

\bibitem[{{Chabrier} and {Baraffe}(1997)}]{ChabrierBaraffe1997}
{Chabrier} G {Baraffe} I (1997) {Structure and evolution of low-mass stars}. A
  \& A 327:1039--1053

\bibitem[{{Chabrier} and {Baraffe}(2000)}]{ChabrierBaraffe2000}
{Chabrier} G {Baraffe} I (2000) {Theory of Low-Mass Stars and Substellar
  Objects}. ARA\&A 38:337--377

\bibitem[{{Comer{\'o}n} et~al.(2000){Comer{\'o}n}, {Neuh{\"a}user}, and
  {Kaas}}]{Comeron2000}
{Comer{\'o}n} F, {Neuh{\"a}user} R {Kaas} AA (2000) {Probing the brown dwarf
  population of the Chamaeleon I star forming region}. A\&A 359:269--288

\bibitem[{{Cushing} et~al.(2011){Cushing}, {Kirkpatrick}, {Gelino}, {Griffith},
  {Skrutskie}, {Mainzer}, {Marsh}, {Beichman}, {Burgasser}, {Prato}, {Simcoe},
  {Marley}, {Saumon}, {Freedman}, {Eisenhardt}, and {Wright}}]{Cushing2011}
{Cushing} MC, {Kirkpatrick} JD, {Gelino} CR et~al. (2011) {The Discovery of Y
  Dwarfs using Data from the Wide-field Infrared Survey Explorer (WISE)}. ApJ
  743:50

\bibitem[{Davies and Davies(2010)}]{Davies2010}
Davies JH Davies DR (2010) Earth's surface heat flux. Solid Earth 1(1):5--24,
  \urlprefix\url{http://www.solid-earth.net/1/5/2010/}

\bibitem[{{Desidera}(1999)}]{Desidera1999}
{Desidera} S (1999) {Properties of Hypothetical Planetary Systems around the
  Brown Dwarf Gliese 229B}. \pasp 111:1529--1538

\bibitem[{{Dittmann} et~al.(2017){Dittmann}, {Irwin}, {Charbonneau}, {Bonfils},
  {Astudillo-Defru}, {Haywood}, {Berta-Thompson}, {Newton}, {Rodriguez},
  {Winters}, {Tan}, {Almenara}, {Bouchy}, {Delfosse}, {Forveille}, {Lovis},
  {Murgas}, {Pepe}, {Santos}, {Udry}, {W{\"u}nsche}, {Esquerdo}, {Latham}, and
  {Dressing}}]{Dittmann2017}
{Dittmann} JA, {Irwin} JM, {Charbonneau} D et~al. (2017) {A temperate rocky
  super-Earth transiting a nearby cool star}. \nat 544:333--336

\bibitem[{{Dressing} and {Charbonneau}(2013)}]{DressingCharbonneau2013}
{Dressing} CD {Charbonneau} D (2013) {The Occurrence Rate of Small Planets
  around Small Stars}. ApJ 767:95

\bibitem[{{Dressing} and {Charbonneau}(2015)}]{DressingCharbonneau2015}
{Dressing} CD {Charbonneau} D (2015) {The Occurrence of Potentially Habitable
  Planets Orbiting M Dwarfs Estimated from the Full Kepler Dataset and an
  Empirical Measurement of the Detection Sensitivity}. \apj 807:45

\bibitem[{{Driscoll} and {Barnes}(2015)}]{DriscollBarnes2015}
{Driscoll} PE {Barnes} R (2015) {Tidal Heating of Earth-like Exoplanets around
  M Stars: Thermal, Magnetic, and Orbital Evolutions}. Astrobiology 15:739--760

\bibitem[{{Eggleton} et~al.(1998){Eggleton}, {Kiseleva}, and {Hut}}]{EKH1998}
{Eggleton} PP, {Kiseleva} LG {Hut} P (1998) {The Equilibrium Tide Model for
  Tidal Friction}. ApJ 499:853--+

\bibitem[{{Gillon} et~al.(2011){Gillon}, {Jehin}, {Magain}, {Chantry},
  {Hutsem{\'e}kers}, {Manfroid}, {Queloz}, and {Udry}}]{Gillon2011}
{Gillon} M, {Jehin} E, {Magain} P et~al. (2011) {TRAPPIST: a robotic telescope
  dedicated to the study of planetary systems}. In: European Physical Journal
  Web of Conferences, European Physical Journal Web of Conferences, vol~11, p
  06002, \doi{10.1051/epjconf/20101106002}

\bibitem[{{Gillon} et~al.(2013){Gillon}, {Jehin}, {Delrez}, {Magain}, {Opitom},
  and {Sohy}}]{Gillon2013}
{Gillon} M, {Jehin} E, {Delrez} L et~al. (2013) {SPECULOOS: Search for
  habitable Planets EClipsing ULtra-cOOl Stars}. In: Protostars and Planets VI
  Posters

\bibitem[{{Gillon} et~al.(2016){Gillon}, {Jehin}, {Lederer}, {Delrez}, {de
  Wit}, {Burdanov}, {Van Grootel}, {Burgasser}, {Triaud}, {Opitom}, {Demory},
  {Sahu}, {Bardalez Gagliuffi}, {Magain}, and {Queloz}}]{Gillon2016}
{Gillon} M, {Jehin} E, {Lederer} SM et~al. (2016) {Temperate Earth-sized
  planets transiting a nearby ultracool dwarf star}. \nat 533:221--224

\bibitem[{{Gillon} et~al.(2017){Gillon}, {Triaud}, {Demory}, {Jehin}, {Agol},
  {Deck}, {Lederer}, {de Wit}, {Ingalls}, {Barkaou}, {Benkhaldoun}, {Bolmont},
  {Burdanov}, {Burgasser}, {Burleigh}, {Carey}, {Copperwheat}, {Delrez},
  {Fernandes}, {Holdsworth}, {Kotze}, {Leconte}, {Magain}, {Queloz}, {Raymond},
  {Sefako}, {Selsis}, {Turbet}, and {Van Grootel}}]{Gillon2017}
{Gillon} M, {Triaud} A, {Demory} BO et~al. (2017) {Seven temperate terrestrial
  planets around the nearby ultracool dwarf star TRAPPIST-1}. \nat 542:456--460

\bibitem[{{Gold} and {Soter}(1969)}]{GoldSoter1969}
{Gold} T {Soter} S (1969) {Atmospheric Tides and the Resonant Rotation of
  Venus}. \icarus 11:356--366

\bibitem[{{Goldreich} and {Soter}(1966)}]{GoldreichSoter1966}
{Goldreich} P {Soter} S (1966) {Q in the Solar System}. Icarus 5:375--389

\bibitem[{{Haas} et~al.(2014){Haas}, {Barclay}, {Batalha}, {Bryson},
  {Caldwell}, {Campbell}, {Coughlin}, {Howell}, {Jenkins}, {Klaus}, {Mullally},
  {Sanderfer}, {Sobeck}, {Still}, {Troeltzsch}, and {Twicken}}]{Haas2014}
{Haas} MR, {Barclay} T, {Batalha} NM et~al. (2014) {The Kepler Mission on Two
  Reaction Wheels is K2}. In: American Astronomical Society Meeting Abstracts
  \#223, American Astronomical Society Meeting Abstracts, vol 223, p 228.01

\bibitem[{{Hansen}(2010)}]{Hansen2010}
{Hansen} BMS (2010) {Calibration of Equilibrium Tide Theory for Extrasolar
  Planet Systems}. ApJ 723:285--299

\bibitem[{{Hayashi} and {Nakano}(1963)}]{Hayashi1963}
{Hayashi} C {Nakano} T (1963) {Evolution of Stars of Small Masses in the
  Pre-Main-Sequence Stages}. Progress of Theoretical Physics 30:460--474

\bibitem[{{He} et~al.(2017){He}, {Triaud}, and {Gillon}}]{He2017}
{He} MY, {Triaud} AHMJ {Gillon} M (2017) {First limits on the occurrence rate
  of short-period planets orbiting brown dwarfs}. \mnras 464:2687--2697

\bibitem[{{Hennebelle} and {Chabrier}(2008)}]{HennebelleChabrier2008}
{Hennebelle} P {Chabrier} G (2008) {Analytical Theory for the Initial Mass
  Function: CO Clumps and Prestellar Cores}. ApJ 684:395--410

\bibitem[{{Henning} and {Hurford}(2014)}]{HenningHurford2014}
{Henning} WG {Hurford} T (2014) {Tidal Heating in Multilayered Terrestrial
  Exoplanets}. \apj 789:30

\bibitem[{{Henry} et~al.(2016){Henry}, {Jao}, {Winters}, {Dieterich}, {Finch},
  {Hambly}, {Ianna}, {McCarthy}, {Riedel}, {Subasavage}, and {RECONS
  Team}}]{Henry2016}
{Henry} TJ, {Jao} WC, {Winters} JG et~al. (2016) {The Census of Objects within
  10 Parsecs}. In: American Astronomical Society Meeting Abstracts, American
  Astronomical Society Meeting Abstracts, vol 227, p 142.01

\bibitem[{{Hut}(1981)}]{Hut1981}
{Hut} P (1981) {Tidal evolution in close binary systems}. A \& A 99:126--140

\bibitem[{{Jackson} et~al.(2008){Jackson}, {Barnes}, and
  {Greenberg}}]{Jackson2008}
{Jackson} B, {Barnes} R {Greenberg} R (2008) {Tidal heating of terrestrial
  extrasolar planets and implications for their habitability}. \mnras
  391:237--245

\bibitem[{{Joshi}(2003)}]{Joshi2003}
{Joshi} M (2003) {Climate Model Studies of Synchronously Rotating Planets}.
  Astrobiology 3:415--427

\bibitem[{{Joshi} and {Haberle}(2012)}]{JoshiHaberle2012}
{Joshi} MM {Haberle} RM (2012) {Suppression of the Water Ice and Snow Albedo
  Feedback on Planets Orbiting Red Dwarf Stars and the Subsequent Widening of
  the Habitable Zone}. Astrobiology 12:3--8

\bibitem[{{Kasting} et~al.(1993){Kasting}, {Whitmire}, and
  {Reynolds}}]{Kasting1993}
{Kasting} JF, {Whitmire} DP {Reynolds} RT (1993) {Habitable Zones around Main
  Sequence Stars}. Icarus 101:108--128

\bibitem[{{Kay} et~al.(2016){Kay}, {Opher}, and {Kornbleuth}}]{Kay2016}
{Kay} C, {Opher} M {Kornbleuth} M (2016) {Probability of CME Impact on
  Exoplanets Orbiting M Dwarfs and Solar-like Stars}. \apj 826:195

\bibitem[{{Kirkpatrick}(2013)}]{Kirkpatrick2013}
{Kirkpatrick} JD (2013) {Cold brown dwarfs with WISE: Y dwarfs and the field
  mass function}. Astronomische Nachrichten 334:26--31

\bibitem[{{Kirkpatrick} et~al.(1999){Kirkpatrick}, {Reid}, {Liebert}, {Cutri},
  {Nelson}, {Beichman}, {Dahn}, {Monet}, {Gizis}, and
  {Skrutskie}}]{Kirkpatrick1999}
{Kirkpatrick} JD, {Reid} IN, {Liebert} J et~al. (1999) {Dwarfs Cooler than
  ``M'': The Definition of Spectral Type ``L'' Using Discoveries from the 2
  Micron All-Sky Survey (2MASS)}. ApJ 519:802--833

\bibitem[{{Kopparapu}(2013)}]{Kopparapu2013}
{Kopparapu} RK (2013) {A Revised Estimate of the Occurrence Rate of Terrestrial
  Planets in the Habitable Zones around Kepler M-dwarfs}. ApJL 767:L8

\bibitem[{{Kumar}(1963)}]{Kumar1963}
{Kumar} SS (1963) {The Structure of Stars of Very Low Mass.} ApJ 137:1121

\bibitem[{{Lammer} et~al.(2003){Lammer}, {Selsis}, {Ribas}, {Guinan}, {Bauer},
  and {Weiss}}]{Lammer2003}
{Lammer} H, {Selsis} F, {Ribas} I et~al. (2003) {Atmospheric Loss of Exoplanets
  Resulting from Stellar X-Ray and Extreme-Ultraviolet Heating}. ApJl
  598:L121--L124

\bibitem[{{Leconte} et~al.(2009){Leconte}, {Baraffe}, {Chabrier}, {Barman}, and
  {Levrard}}]{Leconte2009}
{Leconte} J, {Baraffe} I, {Chabrier} G, {Barman} T {Levrard} B (2009)
  {Structure and evolution of the first CoRoT exoplanets: probing the brown
  dwarf/planet overlapping mass regime}. A \& A 506:385--389

\bibitem[{{Leconte} et~al.(2011{\natexlab{a}}){Leconte}, {Chabrier}, {Baraffe},
  and {Levrard}}]{Leconte2011a}
{Leconte} J, {Chabrier} G, {Baraffe} I {Levrard} B (2011{\natexlab{a}}) {The
  radius anomaly in the planet/brown dwarf overlapping mass regime}. Detection
  and Dynamics of Transiting Exoplanets, St~Michel l'Observatoire, France,
  Edited by F~Bouchy; R~Diaz; C~Moutou; EPJ Web of Conferences, Volume 11,
  id03004 11:3004--+

\bibitem[{{Leconte} et~al.(2011{\natexlab{b}}){Leconte}, {Lai}, and
  {Chabrier}}]{Leconte2011}
{Leconte} J, {Lai} D {Chabrier} G (2011{\natexlab{b}}) {Distorted, nonspherical
  transiting planets: impact on the transit depth and on the radius
  determination}. A \& A 528:A41+

\bibitem[{{Leconte} et~al.(2013){Leconte}, {Forget}, {Charnay}, {Wordsworth},
  {Selsis}, {Millour}, and {Spiga}}]{Leconte2013}
{Leconte} J, {Forget} F, {Charnay} B et~al. (2013) {3D climate modeling of
  close-in land planets: Circulation patterns, climate moist bistability, and
  habitability}. \aap 554:A69

\bibitem[{{Leconte} et~al.(2015){Leconte}, {Wu}, {Menou}, and
  {Murray}}]{Leconte2015}
{Leconte} J, {Wu} H, {Menou} K {Murray} N (2015) {Asynchronous rotation of
  Earth-mass planets in the habitable zone of lower-mass stars}. Science
  347:632--635

\bibitem[{{L{\'o}pez Mart{\'i}} et~al.(2004){L{\'o}pez Mart{\'i}},
  {Eisl{\"o}ffel}, {Scholz}, and {Mundt}}]{LopezMarti2004}
{L{\'o}pez Mart{\'i}} B, {Eisl{\"o}ffel} J, {Scholz} A {Mundt} R (2004) {The
  brown dwarf population in the Chamaeleon I cloud}. A\&A 416:555--576

\bibitem[{{Luger} and {Barnes}(2015)}]{LugerBarnes2015}
{Luger} R {Barnes} R (2015) {Extreme Water Loss and Abiotic O2Buildup on
  Planets Throughout the Habitable Zones of M Dwarfs}. Astrobiology 15:119--143

\bibitem[{{Luhman}(2012)}]{Luhman2012}
{Luhman} KL (2012) {The Formation and Early Evolution of Low-Mass Stars and
  Brown Dwarfs}. \araa 50:65--106

\bibitem[{{Mathis}(2015)}]{Mathis2015}
{Mathis} S (2015) {Variation of tidal dissipation in the convective envelope of
  low-mass stars along their evolution}. \aap 580:L3

\bibitem[{{Mignard}(1979)}]{Mignard1979}
{Mignard} F (1979) {The evolution of the lunar orbit revisited. I}. Moon and
  Planets 20:301--315

\bibitem[{{Mohanty} et~al.(2002){Mohanty}, {Basri}, {Shu}, {Allard}, and
  {Chabrier}}]{Mohanty2002}
{Mohanty} S, {Basri} G, {Shu} F, {Allard} F {Chabrier} G (2002) {Activity in
  Very Cool Stars: Magnetic Dissipation in Late M and L Dwarf Atmospheres}. ApJ
  571:469--486

\bibitem[{{Morata} et~al.(2015){Morata}, {Palau}, {Gonz{\'a}lez}, {de
  Gregorio-Monsalvo}, {Ribas}, {Perger}, {Bouy}, {Barrado}, {Eiroa}, {Bayo},
  {Hu{\'e}lamo}, {Morales-Calder{\'o}n}, and {Rodr{\'{\i}}guez}}]{Morata2015}
{Morata} O, {Palau} A, {Gonz{\'a}lez} RF et~al. (2015) {First Detection of
  Thermal Radiojets in a Sample of Proto-brown Dwarf Candidates}. \apj 807:55

\bibitem[{{Morley} et~al.(2017){Morley}, {Kreidberg}, {Rustamkulov},
  {Robinson}, and {Fortney}}]{Morley2017}
{Morley} CV, {Kreidberg} L, {Rustamkulov} Z, {Robinson} T {Fortney} JJ (2017)
  {Observing the Atmospheres of Known Temperate Earth-sized Planets with JWST}.
  ArXiv e-prints

\bibitem[{{Muirhead} et~al.(2012){Muirhead}, {Johnson}, {Apps}, {Carter},
  {Morton}, {Fabrycky}, {Pineda}, {Bottom}, {Rojas-Ayala}, {Schlawin},
  {Hamren}, {Covey}, {Crepp}, {Stassun}, {Pepper}, {Hebb}, {Kirby}, {Howard},
  {Isaacson}, {Marcy}, {Levitan}, {Diaz-Santos}, {Armus}, and
  {Lloyd}}]{Muirhead2012}
{Muirhead} PS, {Johnson} JA, {Apps} K et~al. (2012) {Characterizing the Cool
  KOIs. III. KOI 961: A Small Star with Large Proper Motion and Three Small
  Planets}. ApJ 747:144

\bibitem[{{Nakajima} et~al.(1995){Nakajima}, {Oppenheimer}, {Kulkarni},
  {Golimowski}, {Matthews}, and {Durrance}}]{Nakajima1995}
{Nakajima} T, {Oppenheimer} BR, {Kulkarni} SR et~al. (1995) {Discovery of a
  cool brown dwarf}. Nature 378:463--465

\bibitem[{{Neron de Surgy} and {Laskar}(1997)}]{DeSurgyLaskar1997}
{Neron de Surgy} O {Laskar} J (1997) {On the long term evolution of the spin of
  the Earth.} A \& A 318:975--989

\bibitem[{{O'Malley-James} and {Kaltenegger}(2017)}]{O'Malley-James2017}
{O'Malley-James} JT {Kaltenegger} L (2017) {UV surface habitability of the
  TRAPPIST-1 system}. \mnras 469:L26--L30

\bibitem[{{Osten} et~al.(2015){Osten}, {Melis}, {Stelzer}, {Bannister},
  {Radigan}, {Burgasser}, {Wolszczan}, and {Luhman}}]{Osten2015}
{Osten} RA, {Melis} C, {Stelzer} B et~al. (2015) {The Deepest Constraints on
  Radio and X-Ray Magnetic Activity in Ultracool Dwarfs from WISE
  J104915.57-531906.1}. \apjl 805:L3

\bibitem[{{Owen} and {Alvarez}(2016)}]{OwenAlvarez2016}
{Owen} JE {Alvarez} MA (2016) {UV Driven Evaporation of Close-in Planets:
  Energy-limited, Recombination-limited, and Photon-limited Flows}. \apj 816:34

\bibitem[{{Padoan} and {Nordlund}(2004)}]{PadoanNordlund2004}
{Padoan} P {Nordlund} {\AA} (2004) {The ``Mysterious'' Origin of Brown Dwarfs}.
  ApJ 617:559--564

\bibitem[{{Phan-Bao} et~al.(2001){Phan-Bao}, {Guibert}, {Crifo}, {Delfosse},
  {Forveille}, {Borsenberger}, {Epchtein}, {Fouqu{\'e}}, and
  {Simon}}]{Phan-Bao2001}
{Phan-Bao} N, {Guibert} J, {Crifo} F et~al. (2001) {New neighbours: IV. 30
  DENIS late-M dwarfs between 15 and 30 parsecs}. A\&A 380:590--598

\bibitem[{{Pizzolato} et~al.(2003){Pizzolato}, {Maggio}, {Micela}, {Sciortino},
  and {Ventura}}]{Pizzolato2003}
{Pizzolato} N, {Maggio} A, {Micela} G, {Sciortino} S {Ventura} P (2003) {The
  stellar activity-rotation relationship revisited: Dependence of saturated and
  non-saturated X-ray emission regimes on stellar mass for late-type dwarfs}.
  A\&A 397:147--157

\bibitem[{{Ranjan} et~al.(2017){Ranjan}, {Wordsworth}, and
  {Sasselov}}]{Ranjan2017}
{Ranjan} S, {Wordsworth} R {Sasselov} DD (2017) {The Surface UV Environment on
  Planets Orbiting M Dwarfs: Implications for Prebiotic Chemistry and the Need
  for Experimental Follow-up}. \apj 843:110

\bibitem[{{Rauer} et~al.(2011){Rauer}, {Gebauer}, {Paris}, {Cabrera}, {Godolt},
  {Grenfell}, {Belu}, {Selsis}, {Hedelt}, and {Schreier}}]{Rauer2011}
{Rauer} H, {Gebauer} S, {Paris} PV et~al. (2011) {Potential biosignatures in
  super-Earth atmospheres. I. Spectral appearance of super-Earths around M
  dwarfs}. \aap 529:A8

\bibitem[{{Rebolo} et~al.(1995){Rebolo}, {Zapatero Osorio}, and
  {Mart{\'{\i}}n}}]{Rebolo1995}
{Rebolo} R, {Zapatero Osorio} MR {Mart{\'{\i}}n} EL (1995) {Discovery of a
  brown dwarf in the Pleiades star cluster}. Nature 377:129--131

\bibitem[{{Ribas} et~al.(2016){Ribas}, {Bolmont}, {Selsis}, {Reiners},
  {Leconte}, {Raymond}, {Engle}, {Guinan}, {Morin}, {Turbet}, {Forget}, and
  {Anglada-Escud{\'e}}}]{Ribas2016}
{Ribas} I, {Bolmont} E, {Selsis} F et~al. (2016) {The habitability of Proxima
  Centauri b. I. Irradiation, rotation and volatile inventory from formation to
  the present}. \aap 596:A111

\bibitem[{{Rodler} and {L{\'o}pez-Morales}(2014)}]{Rodler2014}
{Rodler} F {L{\'o}pez-Morales} M (2014) {Feasibility Studies for the Detection
  of O$_{2}$ in an Earth-like Exoplanet}. \apj 781:54

\bibitem[{{Rutledge} et~al.(2000){Rutledge}, {Basri}, {Mart{\'{\i}}n}, and
  {Bildsten}}]{Rutledge2000}
{Rutledge} RE, {Basri} G, {Mart{\'{\i}}n} EL {Bildsten} L (2000) {Chandra
  Detection of an X-Ray Flare from the Brown Dwarf LP 944-20}. \apjl
  538:L141--L144

\bibitem[{{Salpeter}(1955)}]{Salpeter1955}
{Salpeter} EE (1955) {The Luminosity Function and Stellar Evolution.} \apj
  121:161

\bibitem[{{Segura} et~al.(2005){Segura}, {Kasting}, {Meadows}, {Cohen},
  {Scalo}, {Crisp}, {Butler}, and {Tinetti}}]{Segura2005}
{Segura} A, {Kasting} JF, {Meadows} V et~al. (2005) {Biosignatures from
  Earth-Like Planets Around M Dwarfs}. Astrobiology 5:706--725

\bibitem[{{Segura} et~al.(2010){Segura}, {Walkowicz}, {Meadows}, {Kasting}, and
  {Hawley}}]{Segura2010}
{Segura} A, {Walkowicz} LM, {Meadows} V, {Kasting} J {Hawley} S (2010) {The
  Effect of a Strong Stellar Flare on the Atmospheric Chemistry of an
  Earth-like Planet Orbiting an M Dwarf}. Astrobiology 10:751--771

\bibitem[{{Selsis} et~al.(2007){Selsis}, {Kasting}, {Levrard}, {Paillet},
  {Ribas}, and {Delfosse}}]{Selsis2007}
{Selsis} F, {Kasting} JF, {Levrard} B et~al. (2007) {Habitable planets around
  the star Gliese 581?} A \& A 476:1373--1387

\bibitem[{{Showman} and {Polvani}(2011)}]{Showman2011}
{Showman} AP {Polvani} LM (2011) {Equatorial Superrotation on Tidally Locked
  Exoplanets}. \apj 738:71

\bibitem[{{Skrutskie} et~al.(2006){Skrutskie}, {Cutri}, {Stiening}, {Weinberg},
  {Schneider}, {Carpenter}, {Beichman}, {Capps}, {Chester}, {Elias}, {Huchra},
  {Liebert}, {Lonsdale}, {Monet}, {Price}, {Seitzer}, {Jarrett}, {Kirkpatrick},
  {Gizis}, {Howard}, {Evans}, {Fowler}, {Fullmer}, {Hurt}, {Light}, {Kopan},
  {Marsh}, {McCallon}, {Tam}, {Van Dyk}, and {Wheelock}}]{Skrutskie2006}
{Skrutskie} MF, {Cutri} RM, {Stiening} R et~al. (2006) {The Two Micron All Sky
  Survey (2MASS)}. AJ 131:1163--1183

\bibitem[{{Spencer} et~al.(2000){Spencer}, {Jessup}, {McGrath}, {Ballester},
  and {Yelle}}]{Spencer2000}
{Spencer} JR, {Jessup} KL, {McGrath} MA, {Ballester} GE {Yelle} R (2000)
  {Discovery of Gaseous S$_{2}$ in Io's Pele Plume}. Science 288:1208--1210

\bibitem[{{Spencer} et~al.(2007){Spencer}, {Stern}, {Cheng}, {Weaver},
  {Reuter}, {Retherford}, {Lunsford}, {Moore}, {Abramov}, {Lopes}, {Perry},
  {Kamp}, {Showalter}, {Jessup}, {Marchis}, {Schenk}, and
  {Dumas}}]{Spencer2007}
{Spencer} JR, {Stern} SA, {Cheng} AF et~al. (2007) {Io Volcanism Seen by New
  Horizons: A Major Eruption of the Tvashtar Volcano}. Science 318:240

\bibitem[{{Spiegel} et~al.(2011){Spiegel}, {Burrows}, and
  {Milsom}}]{Spiegel2011}
{Spiegel} DS, {Burrows} A {Milsom} JA (2011) {The Deuterium-burning Mass Limit
  for Brown Dwarfs and Giant Planets}. ApJ 727:57--+

\bibitem[{{Tabataba-Vakili} et~al.(2016){Tabataba-Vakili}, {Grenfell},
  {Grie{\ss}meier}, and {Rauer}}]{Tabataba-Vakili2016}
{Tabataba-Vakili} F, {Grenfell} JL, {Grie{\ss}meier} JM {Rauer} H (2016)
  {Atmospheric effects of stellar cosmic rays on Earth-like exoplanets orbiting
  M-dwarfs}. \aap 585:A96

\bibitem[{{Triaud} et~al.(2013){Triaud}, {Gillon}, {Selsis}, {Winn}, {Demory},
  {Artigau}, {Laughlin}, {Seager}, {Helling}, {Mayor}, {Albert}, {Anderson},
  {Bolmont}, {Doyon}, {Forveille}, {Hagelberg}, {Leconte}, {Lendl},
  {Littlefair}, {Raymond}, and {Sahlmann}}]{Triaud2013}
{Triaud} AHMJ, {Gillon} M, {Selsis} F et~al. (2013) {A search for rocky planets
  transiting brown dwarfs}. ArXiv e-prints

\bibitem[{{Turbet} et~al.(2016){Turbet}, {Leconte}, {Selsis}, {Bolmont},
  {Forget}, {Ribas}, {Raymond}, and {Anglada-Escud{\'e}}}]{Turbet2016}
{Turbet} M, {Leconte} J, {Selsis} F et~al. (2016) {The habitability of Proxima
  Centauri b. II. Possible climates and observability}. \aap 596:A112

\bibitem[{{Turbet} et~al.(2017){Turbet}, {Bolmont}, {Leconte}, {Forget},
  {Selsis}, {Tobie}, {Caldas}, {Naar}, and {Gillon}}]{Turbet2017}
{Turbet} M, {Bolmont} E, {Leconte} J et~al. (2017) {Climate diversity on cool
  planets around cool stars with a versatile 3-D Global Climate Model: the case
  of TRAPPIST-1 planets}. ArXiv e-prints

\bibitem[{{Venot} et~al.(2016){Venot}, {Rocchetto}, {Carl}, {Roshni Hashim},
  and {Decin}}]{Venot2016}
{Venot} O, {Rocchetto} M, {Carl} S, {Roshni Hashim} A {Decin} L (2016)
  {Influence of Stellar Flares on the Chemical Composition of Exoplanets and
  Spectra}. \apj 830:77

\bibitem[{{Vida} et~al.(2017){Vida}, {K{\H o}v{\'a}ri}, {P{\'a}l}, {Ol{\'a}h},
  and {Kriskovics}}]{Vida2017}
{Vida} K, {K{\H o}v{\'a}ri} Z, {P{\'a}l} A, {Ol{\'a}h} K {Kriskovics} L (2017)
  {Frequent Flaring in the TRAPPIST-1 System -- Unsuited for Life?} \apj
  841:124

\bibitem[{{Watson} et~al.(1981){Watson}, {Donahue}, and {Walker}}]{Watson1981}
{Watson} AJ, {Donahue} TM {Walker} JCG (1981) {The dynamics of a rapidly
  escaping atmosphere - Applications to the evolution of earth and Venus}.
  \icarus 48:150--166

\bibitem[{{Williams} et~al.(2014){Williams}, {Cook}, and
  {Berger}}]{WilliamsCookBerger2014}
{Williams} PKG, {Cook} BA {Berger} E (2014) {Trends in Ultracool Dwarf
  Magnetism. I. X-Ray Suppression and Radio Enhancement}. \apj 785:9

\bibitem[{{Wordsworth} et~al.(2011){Wordsworth}, {Forget}, {Selsis}, {Millour},
  {Charnay}, and {Madeleine}}]{Wordsworth2011}
{Wordsworth} RD, {Forget} F, {Selsis} F et~al. (2011) {Gliese 581d is the First
  Discovered Terrestrial-mass Exoplanet in the Habitable Zone}. ApJ Lett
  733:L48+

\bibitem[{{Zapatero Osorio} et~al.(1997){Zapatero Osorio}, {Rebolo}, {Martin},
  {Basri}, {Magazzu}, {Hodgkin}, {Jameson}, and {Cossburn}}]{Zapatero1997}
{Zapatero Osorio} MR, {Rebolo} R, {Martin} EL et~al. (1997) {New Brown Dwarfs
  in the Pleiades Cluster}. ApJ Lett 491:L81

\end{thebibliography}

\end{document}